\documentstyle[psfig,onecolumn,referee]{mn2e}

\input{epsf}

\newif\ifAMStwofonts
\AMStwofontstrue

\newcommand{\go}{\mathrel{\raise.3ex\hbox{$>$}\mkern-14mu
             \lower0.6ex\hbox{$\sim$}}}
\newcommand{\lo}{\mathrel{\raise.3ex\hbox{$<$}\mkern-14mu
             \lower0.6ex\hbox{$\sim$}}}
\newcommand{\lp}{\left(}
\newcommand{\rp}{\right)}
\newcommand{\rd}{{\rm d}}
\newcommand{\lb}{\left[}
\newcommand{\rb}{\right]}

\newcommand{\vecr}{\bmath r}
\newcommand{\vecB}{\bmath B}
\newcommand{\vecH}{\bmath H}
\newcommand{\vechatB}{\hat{\bmath B}}
\newcommand{\vecmu}{\bmath \mu}
\newcommand{\vecmui}{{\bmath \mu_{\rm i}}}
\newcommand{\mui}{{\mu_{\rm i}}}
\newcommand{\ri}{{r_{\rm i}}}

\newcommand{\vecW}{\bmath \Omega}
\newcommand{\vechatW}{\hat{\bmath \Omega}}
\newcommand{\vecE}{\bmath E}
\newcommand{\vecD}{\bmath D}

\newcommand{\veck}{{\bmath k}}
\newcommand{\vechatk}{\hat{\bmath k}}

\newcommand{\vechaty}{\hat{\bmath y}}

\newcommand{\vechatz}{\hat{\bmath z}}

\newcommand{\vechatX}{\hat{\bmath X}}
\newcommand{\vechatY}{\hat{\bmath Y}}
\newcommand{\vechatZ}{\hat{\bmath Z}}

\newcommand{\dt}{{\bmath{\bepsilon}}}

\newcommand{\singhsq}{\sin^2\theta_B}
\newcommand{\cosgh}{\cos\theta_B}
\newcommand{\cosghsq}{\cos^2\theta_B}
\newcommand{\ggtheta}{\lp1-\beta\cosgh\rp}

\newcommand{\thetab}{\theta_B}

\newcommand{\me}{m_{\rm e}}
\newcommand{\Bq}{B_{\rm Q}}

\newcommand{\Ps}{P_1}

\newcommand{\FQb}{\bar{F_Q}}
 
\title[]{Polarization Evolution in A Strongly Magnetized Vacuum:
QED Effect and Polarized X-ray Emission from Magnetized Neutron Stars}
\author[C. Wang and D. Lai]{Chen Wang$^{1,2}$ and Dong Lai$^{1}$ \\
$^{1}$ Department of Astronomy, Cornell University, Ithaca, NY 14853, USA \\
$^{2}$ National Astronomical Observatories, Chinese Academy of Sciences.
A20 Datun Road, Chaoyang District, Beijing 100012, China \\
{\rm E-mail: cwang, dong@astro.cornell.edu}}
\date{Accepted 2009 xxx,
      Received 2009 xxx;
      in original form 2009 xxx}

\pagerange{\pageref{firstpage}--\pageref{lastpage}}
\begin{document}
\maketitle

\label{firstpage}

\begin{abstract}
X-ray photons emitted from the surface or atmosphere of a magnetized
neutron star are highly polarized. However, the observed polarization
may be modified due to photon propagation through the star's
magnetosphere. For photon frequencies much larger than the typical
radio frequency, vacuum birefringence due to strong-field quantum
electrodynamics dominates over the plasma effect.  We study the
evolution of photon polarization in the magnetized QED vacuum of a
neutron star magnetosphere, paying particular attention to the
propagation effect across the quasi-tangential (QT) point, where the
photon momentum is nearly aligned with the magnetic field. In
agreement with previous studies, we find that in most regions of the
magnetosphere, the photon polarization modes are decoupled due to
vacuum birefringence, and therefore a large net linear polarization
can be expected when the radiation escapes the magnetosphere.
However, we show that the X-ray polarization may change significantly
when the photon passes through the QT region. When averaging over a
finite emission area, the net effect of QT propagation is to reduce
the degree of linear polarization; the reduction factor depends on the
photon energy, magnetic field strength, geometry, rotation phase and
the emission area, and can be more than a factor of two. We derive the
general conditions under which the QT propagation effect is important,
and provide an easy-to-use prescription to account for the QT effect
for most practical calculations of X-ray polarization signals from
magnetic neutron stars. For a neutron star with a dipole magnetic
field, the QT effect can be important for emission from the polar cap
for certain magnetic field and energy ranges, and is negligible for
emission from the entire stellar surface.

\end{abstract}


\setcounter{equation}{0}
\section{Introduction}

Thermal, surface emission from neutron stars (NSs) has the potential
of providing invaluable information on the physical properties and
evolution of NSs (equation of state at super-nuclear densities,
cooling history, magnetic field, surface composition, different
populations; see, e.g., Yakovlev \& Pethick 2004; Harding \& Lai
2006).  With X-ray telescope such as {\it Chandra} and {\it
  XMM-Newton}, the last decade has seen significant observational
progress, revealing the surface magnetic field geometry of isolated
pulsars with phase-resolved spectroscopy, and constraining the cooling
physics from thermal emission of young NSs in supernova remnants
(e.g., Kaspi et al.~2006).  In addition, thermal emission from seven
isolated, radio-quiet NSs has been studied in detail, revealing
absorption features in their spectra in many cases (see, e.g., van
Kerkwijk \& Kaplan 2007; Kaplan 2008).

It has been recognized that in addition to imaging, timing and
spectroscopy, X-ray polarimetry provides a new way to study many
high-energy astrophysical sources, particularly magnetic NSs. Recent
advances in detector technology suggest that polarimetry study of
X-ray sources holds great promise in the future (e.g., Costa et
al.~2008; Swank et al.~2008).

The surface emission from magnetized NSs (with $B\go 10^{12}$\,G) is
highly polarized (e.g., Gnedin \& Sunyaev 1974; Meszaros et al.~1988;
Pavlov \& Zavlin 2000) for the following reason.  In the magnetized
plasma that characterizes NS atmospheres, X-ray photons propagate in
two normal modes: the ordinary mode (O-mode, or $\parallel$-mode) is
mostly polarized parallel to the $\veck$-$\vecB$ plane, while the
extraordinary mode (X-mode, or $\perp$-mode) is mostly polarized
perpendicular to the $\veck$-$\vecB$ plane, where $\veck$ is the
photon wave vector and $\vecB$ is the external magnetic field (e.g.,
Meszaros 1992).  This description of normal modes applies under
typical conditions, when the photon energy $E$ is much less than the
electron cyclotron energy $E_{Be}=\hbar eB/(m_ec)=11.6\,B_{12}$\,keV
[where $B_{12}=B/(10^{12}\,{\rm G})$], $E$ is not too close to the ion
cyclotron energy $E_{Bi}=6.3\,B_{12}(Z/A)$\,eV (where $Z$ and $A$ are
the charge number and mass number of the ion), the plasma density is
not too close to the vacuum resonance (see below) and $\theta_B$ (the
angle between $\veck$ and $\vecB$) is not close to zero.  Under these
conditions, the X-mode opacity (due to scattering and absorption) is
greatly suppressed compared to the O-mode opacity, $\kappa_X\sim
(E/E_{Be})^2\kappa_O$ (e.g. Lodenquai et al.~1974; Meszaros 1992;
Potekhin \& Chabrier 2003). As a result, the X-mode photons escape
from deeper, hotter layers of the NS atmosphere than the O-mode
photons, and the emergent radiation is linearly polarized to a high
degree (e.g., Pavlov \& Zavlin 2000; Ho \& Lai 2001,2003; van
Adelsberg \& Lai 2006).  Measurements of X-ray polarization,
particularly when phase-resolved and measured in different energy
bands, could provide unique constraints on the NS magnetic field
strength and geometry.

It has long been predicted from quantum electrodynamics (QED) that in
a strong magnetic field the vacuum becomes birefringent (e.g.,
Schwinger 1951; Adler 1971). While this vacuum polarization effect
makes the photon index of refraction deviate from unity only when
$B\go 300 B_{\rm Q}$, where $B_{\rm Q}=m_e^2c^3/(e\hbar)=4.414\times
10^{13}$\,G is the critical QED field strength, it can significantly
affect the spectra of polarization signals from magnetic NSs in more
subtle way, at much lower field strengths (see section 2 of Lai \& Ho
2003a for a qualitative explanation). In particular, the combined
effects of vacuum polarization and magnetized plasma gives rise to a
``vacuum resonance'', at which the contributions from these two
effects (plasma and vacuum polarization) to the dielectric tensor
``compensate'' each other (Gnedin et al.~1978; Meszaros \& Ventura
1979; Pavlov \& Gnedin 1984; Lai \& Ho 2002).  A photon may convert
from the high-opacity mode to the low-opacity one and vice verse when
it crosses the vacuum resonance region in the inhomogeneous NS
atmosphere. For $B\go 7\times 10^{14}$\,G, this vacuum resonance
phenomenon tends to soften the hard spectral tail due to the
non-greyness of the atmospheric opacities and suppress the width of
absorption lines, while for $B\lo 7\times 10^{14}$\,G, the spectrum is
unaffected (see Lai \& Ho 2002,2003a; Ho \& Lai 2003; van Adelsberg \&
Lai 2006).

The QED-induced vacuum birefringence influences the X-ray polarization
signals from magnetic NSs in two different ways.  (i) {\it Photon mode
  conversion in the NS atmosphere:} Since the mode conversion depends
on photon energy and magnetic field strength, this vacuum resonance
effect gives rise to a unique energy-dependent polarization signal in
X-rays: For ``normal'' field strengths ($B\lo 7\times 10^{13}$\,G),
the plane of linear polarization at the photon energy $E\lo 1$\,keV is
perpendicular to that at $E\go 4$\,keV, while for ``superstrong''
field strengths ($B\go 7\times 10^{13}$\,G), the polarization planes
at different energies coincide (Lai \& Ho 2003b; van Adelsberg \& Lai
2006).  (ii) {\it Polarization mode decoupling in the magnetosphere:}
The birefringence of the magnetized QED vacuum decouples the photon
polarization modes, so that as a polarized photon leaves the NS
surface and propagates through the magnetosphere, its polarization
direction follows the direction of the magnetic field up to a large
radius (the so-called polarization limiting radius). The result is
that although the magnetic field orientations over the NS surface may
vary widely, the polarization directions of the photon originating
from different surface regions tend to align, giving rise to large
observed polarization signals (Heyl \& Shaviv 2002; van Adelsberg \&
Lai 2006).

In this paper we examine in detail the photon polarization evolution
in the magnetized QED vacuum of a NS magnetosphere [Effect (ii) in the
  last paragraph]. We are particularly interested in photon
propagation across the {\it Quasi-tangential} Region (QT region): As
the photon travels through the magnetosphere, it may cross the region
where its wave vector is aligned or nearly aligned with the magnetic
field (i.e., $\theta_B$ is zero or small). In such a QT region, the
two photon modes ($\parallel$ and $\perp$ modes) become (nearly)
identical, and mode coupling may occur, thereby affecting the
polarization alignment.  Our previous analytical (or semi-analytical)
treatment of the alignment effect (Lai \& Ho 2003b; van Adelsberg \&
Lai 2006) focused on the region far away from the NS surface, thus did
not not include the QT region (which typically lies within a few
stellar radii). The numerical ray integration presented in Heyl \&
Shaviv (2002; see also Heyl et al.~2003) should in principle have
included such QT region, but no systematic characterization of the QT
propagation effect on the final photon polarization was attempted
there. As we show in this paper, polarization evolution through the QT
region is sufficiently subtle (e.g. the effect varies on small length
scales across the emission region) that a careful examination of its
effect is necessary. The purpose of this paper is to study the
evolution of high-energy (X-ray) photon polarization in NS
magnetospheres and to quantitatively assess the QT propagation effect.

The remainder of our paper is organized as follows: Section 2
summarizes the basic equations for studying photon polarization
evolution in magnetized QED vacuum.  In section 3 we examine the
general behavior of the polarization evolution across a QT region in a
generic magnetic geometry. In section 4 we present detailed
calculations in the case of dipole magnetic field and consider
emissions from both the polar cap and the other regions of the NS
surface. We provide a simple prescription (see section 4.3) for
including the QT effect in the calculations of the observed
polarization signals. In section 5 we discuss the implications of our
results for the X-ray polarization signals from magnetic NSs and the
prospect of using X-ray polarimetry to probe strong-field QED.


\section{Polarization Evolution in Highly Magnetized QED Vacuum: Equations}
\label{sec:wee}

The magnetospheres of pulsars and magnetars consist of relativistic
electron-positron pairs streaming along magnetic field lines.  The
Lorentz factor $\gamma$ of the streaming motion and the plasma density
$N$ are uncertain.  For the open field line region of radio pulsars,
pair cascade simulations generally give $\gamma\sim 10^2-10^4$ and
$\eta\equiv N/N_{\rm GJ}\sim 10^2-10^5$ (e.g., Daugherty \& Harding
1982; Hibschman \& Arons 2001; Medin \& Lai 2009), while recent
theoretical works suggest that the corona of magnetars consist of pair
plasma with $\gamma$ up to $10^3$ and $\eta\sim 2\times
10^3(R_\ast/r)$ (where $R_\ast$ is the stellar radius; Thompson et
al.~2002; Beloborodov \& Thompson 2006), where $N_{\rm GJ} =(\Omega
B)/(2\pi ec)$
is the Goldreich-Julian density. In general, both the plasma and vacuum
polarization affect the photon modes in the magnetosphere. For a
given photon energy $E$, $\gamma$ and $B$, the vacuum resonance occurs
at the density (Wang \& Lai 2007)
\begin{equation}
N_{\rm V}=5.80\times 10^{30} B_{13}^2E_1^2\gamma_3^3
           \ggtheta^2F(b)\,{\rm cm}^{-3},
\label{eqrhores}
\end{equation}
where $\beta=\sqrt{1-1/\gamma^2}$, $\theta_B$ is $\veck$-$\vecB$
angle, $E_1=E/(1\,{\rm keV})$, $B_{13}=B/(10^{13}\,{\rm G})$,
$\gamma_3=\gamma/10^3$, $F(b)$ is equal to unity for $b=B/B_{\rm Q}\ll 1$
and is at most of order a few for $B\lo 10^{15}$\,G. For the typical
photon energy of interest in this paper, the magnetosphere plasma
density $N$ is much less than $N_{\rm V}$, and vacuum birefringence
dominates over the plasma effect. To put it in another way, at given
$B$, $\gamma$ and density $N$ (or $\eta$), we can define the vacuum
resonance photon energy:
\begin{equation}
E_{\rm V} = 3.48\times10^{-10}\lb \Ps^{-1}B_{13}^{-1}\eta
          \gamma_3^{-3}\ggtheta^{-2}F^{-1}\rb^{1/2}\,{\rm keV}.
\label{eq:nuV}
\end{equation}
where $\Ps$ is the NS spin period in units of 1 second. Throughout
this paper, we shall be interested in photon energies $E \gg E_{\rm
  V}$, so that the wave modes are determined by the vacuum
polarization effect.

In this section we summarize the key equations for studying the
polarization evolution of X-rays in NS magnetospheres.

\subsection{Wave Modes}

The dielectric tensor and the inverse permeability tensor of a
magnetized QED vacuum take the form:
\begin{equation}
\dt=a{\bmath I}+q\vechatB\vechatB,
\quad
\bmath{\mu^{-1}}=a{\bmath I}+m\vechatB\vechatB,
\label{eq:dt}
\end{equation}
where $\bmath I$ is the unit tensor and $\vechatB=\vecB/B$ is the unit
vector along $\vecB$.  In the low frequency limit, $E=\hbar\omega\ll\me
c^2$, the general expressions for the vacuum polarization coefficients
$a$, $q$, and $m$ are given in Adler (1971) and Heyl \&
Hernquist~(1997).  For $B\ll\Bq=m_e^2c^3/(e\hbar)=4.414\times
10^{13}$\,G, they are given by
\begin{equation}
a=1-2\delta_{\rm V}, \quad q=7\delta_{\rm V}, \quad m=-4\delta_{\rm V},
\end{equation}
with
\begin{equation}
\delta_{\rm V}=\frac{\alpha_{\rm F}}{45\pi}\lp\frac{B}{B_{\rm Q}}\rp^2
\simeq 2.65\times10^{-6}B_{13}^2.
\end{equation}
Here $\alpha_{\rm F}=e^2/\hbar c=1/137$ is the fine structure constant.  For
$B\gg\Bq$, simple expressions for $a$, $q$, $m$ are given in Ho \&
Lai~(2003) (see also Potekhin et al. 2004 for general fitting
formulae).

Using the relations $\vecD=\dt\cdot\vecE$,
$\vecB=\bmath{\mu}\cdot\vecH$ and the Maxwell equations, we obtain the
equation for plane waves with $\vecE\propto
e^{i(\veck\cdot\vecr-\omega t)}$
\begin{equation}
\left\{ \frac{1}{a}\epsilon_{ij}+n^2\lb\hat{k}_i\hat{k}_j-\delta_{ij}
 - \frac{m}{a}(\hat{k}\times\hat{B})_i(\hat{k}\times\hat{B})_j
 \rb \right\} E_j=0, \label{eq:nrefract}
\end{equation}
where $n=ck/\omega$ is the refractive index and $\vechatk=\veck/k$.
The dielectric tensor are given by eq.~(\ref{eq:dt}).  In the
coordinate system $xyz$ (with $\veck$ along the $z$-axis and $\vecB$
in the $x$-$z$ plane, such that
$\vechatk\times\vechatB=-\sin\thetab\vechaty$), we can solve
equation~(\ref{eq:nrefract}) to obtain the two eigenmodes: the
$\parallel$-mode (or ordinary mode, polarized in the $\veck$-$\vecB$
plane) and $\perp$-modes (or extraordinary mode, polarized
perpendicular to the $\veck$-$\vecB$ plane).  The refractive indices
and polarization states of these two modes are
\begin{eqnarray}
n_\parallel=\lp\frac{a+q}{a+q\cosghsq}\rp^{1/2},
&&\quad |E_x/E_y|_\parallel=\infty; \nonumber\\
n_\perp=\lp\frac{a}{a+m\singhsq}\rp^{1/2}
&&\quad |E_y/E_x|_\perp=\infty.  \label{eq:n}
\end{eqnarray}
The difference between the two refractive indices is
(for $q,|m|\ll 1$)
\begin{equation}
\Delta n=n_{\parallel}-n_{\perp}\simeq \frac{1}{2}(q+m)\singhsq.\label{eq:dn}
\end{equation}

\subsection{Mode Evolution Equations and the Adiabatic Condition}
\label{sec:ad}

A general polarized electromagnetic wave with frequency $\omega$
traveling in the $z$-direction can be written as a superposition of
the two modes:
\begin{equation}
\vecE(z)=A_{\parallel}(z)\vecE_{\parallel}(z)+A_{\perp}(z)\vecE_{\perp}(z),
\label{field}
\end{equation}
Note that both $A_\parallel$, $A_\perp$ and $\vecE_\parallel$,
$E_\perp$ depend on $z$.  Substituting equation~(\ref{field}) into the
wave equation
\begin{equation}
\nabla\times\lp\bmu^{-1}\cdot\nabla\times\vecE\rp
  =\frac{\omega^2}{c^2} \dt\cdot\vecE,
\end{equation}
we obtain the amplitude evolution equations (see Adelsberg \& Lai
2006)
\begin{equation}
i\frac{\rd}{\rd z}\left(\begin{array}{c}A_\parallel\\ A_\perp\end{array}\right)
\simeq \left(\begin{array}{cc}
-(\omega/c)\Delta n/2 & i\varphi_B'\\
-i\varphi_B' & (\omega/c)\Delta n/2
\end{array} \right)
\left(\begin{array}{c}A_\parallel\\ A_\perp\end{array}\right),
\label{eq:me}
\end{equation}
where $\varphi_B'=\rd\varphi_B/\rd z$. Here $\varphi_B$ is the azimuthal
angle of $\vecB$ in the fixed frame $XYZ$ with the $Z$-axis parallel
to the line of sight $\vechatk$ (Note that the $xyz$-frame introduced
in section 2.1 rotates around the $Z$-axis since $\vecB$ changes along
the ray).  In deriving equation~(\ref{eq:me}), we have used the
geometric optical approximation $|dA/dz|\ll k|A|$.

The condition for the adiabatic evolution of wave modes is
\begin{equation}
\Gamma_{\rm ad}=\left|\frac{\Delta n\omega}{2\varphi_B'c}\right|\gg1.
\label{eq:Gamma_ad}
\end{equation}
Here $\Gamma_{\rm ad}$ is the adiabaticity parameter which changes
along the photon ray. In the adiabatic limit, the photon modes are
{\it decoupled}, and the photon will always stay in the initial mode
although the mode itself changes along the ray following the direction
of $\vecB_\perp$ (the projection of $\vecB$ in the XY
plane). Substituting eq.~(\ref{eq:dn}) into eq.~(\ref{eq:Gamma_ad}),
we have
\begin{equation}
\Gamma_{\rm ad}\simeq \left|\frac{(q+m)\omega}{4c}
 \frac{\singhsq}{\varphi_B'}\right|
\simeq 1.0\times10^7E_1B_{13}^2\frac{\singhsq}{|\varphi_B'|}F(b),
\label{eq:Gamma_ad2}
\end{equation}
where
\begin{equation}
F(b)=\frac{q+m}{\alpha_{\rm F}^2(B/B_{\rm Q})^2/(15\pi)}
\end{equation}
is equal to unity for $b=B/B_{\rm Q}\ll 1$ and is at most of order a
few for $B\lo 10^{15}$\,G (see Fig.~1 of Ho \& Lai 2003) --- we will
use $F\simeq 1$ hereafter. The unit of $\varphi_B'$ is km$^{-1}$ and
the factor $\singhsq/\varphi_B'$ is determined by magnetic field
geometry along the ray.

Once the mode amplitudes $A_\parallel$ and $A_\perp$ are known, we can obtain the
Stokes parameters in the fixed $XYZ$-frame from (see van Adelsberg \& Lai
2006)
\begin{eqnarray}
I &=& |A_\parallel|^2+|A_\perp|^2, \nonumber\\
Q &=& \cos2\varphi_B(|A_\parallel|^2-|A_\perp|^2)
      -2\sin2\varphi_B\Re e(A_\parallel A_\perp^*), \nonumber\\
U &=& \sin2\varphi_B(|A_\parallel|^2-|A_\perp|^2)
      +2\cos2\varphi_B\Re e(A_\parallel A_\perp^*), \nonumber\\
V &=& 2\Im m(A_\parallel A_\perp^*).
\end{eqnarray}
Alternatively, we can use equation (\ref{eq:me}) to obtain the
evolution equation for the Stokes parameters (see van Adelsberg \& Lai
2006):
\begin{eqnarray}
\rd Q/\rd Z&=& (\omega/c)\Delta n V \sin2\varphi_B, \nonumber\\
\rd U/\rd Z&=& -(\omega/c)\Delta n V\cos2\varphi_B, \nonumber\\
\rd V/\rd Z&=& -(\omega/c)\Delta n(Q\sin2\varphi_B-U\cos2\varphi_B).
\end{eqnarray}
and $\rd I/\rd Z=0$.

\section{Polarization Evolution Across the Quasi-Tangential Region}
\label{sec:qte}

Here we examine the general behavior of the polarization evolution
when a photon crosses the quasi-tangential (QT) region.  An exact
tangential point is where the photon wave vector $\veck$ is aligned
with $\vecB$, or $\theta_B=0$. Across the tangential point, the
azimuth of the magnetic field, $\varphi_B$, changes by $180^o$.  For a
general magnetic field geometry, not all photon rays have such an
exact tangential point. However, there exits a {\it quasi-tangential}
point where the $\veck$-$\vecB$ angle $\theta_B$ reaches a local
minimum.  Without loss of generality, the magnetic field around the QT
point can be modeled as
\begin{equation}
B_X=\frac{B}{\cal R}s, \quad B_Y=\epsilon B,  \label{eq:simplemode}
\end{equation}
in the fixed $XYZ$-frame with $\vechatZ\parallel\vechatk$.  Here $B$
is the magnitude of $\vecB$, ${\cal R}$ is the curvature radius of the
projected magnetic field line in $XZ$-plane, $s$ measures the distance
along the $Z$-axis (with the QT point at $s=0$). The polar angle and
azimuthal angle of $\vecB$ are given by
\begin{equation}
\sin\theta_B=\frac{B_\perp}{B}=\sqrt{\left(\frac{s}{\cal R}\right)^2+\epsilon^2},
\quad \tan\varphi_B=\frac{B_Y}{B_X}=\frac{\epsilon \cal R}{s}.
\end{equation}
The minimum $\theta_B$ occurs at the QT point
\begin{equation}
\sin\theta_{\rm t}=\sin\theta_B\mid_{s=0}=\epsilon.
\end{equation}
We also have
\begin{equation}
\frac{\rd\varphi_B}{\rd s}=\cos^2\varphi_B\frac{\rd}{\rd s}\left(\frac{B_Y}{B_X}\right)
=-\frac{1}{\epsilon \cal R}\left(1+\frac{s^2}{\epsilon^2{\cal R}^2}\right)^{-1}.
\end{equation}
Thus the adiabaticity parameter along the ray is
\begin{equation}
\Gamma_{\rm ad}=1.0\times10^8E_1B_{13}^2\epsilon^3{\cal R}_1
              \left(1+\frac{s^2}{\epsilon^2{\cal R}^2}\right)^{2},
\label{eq:Gamma_ad_gm}
\end{equation}
where ${\cal R}_1={\cal R}/(10~{\rm km})$.

\begin{figure*}
\begin{tabular}{c}
\psfig{figure=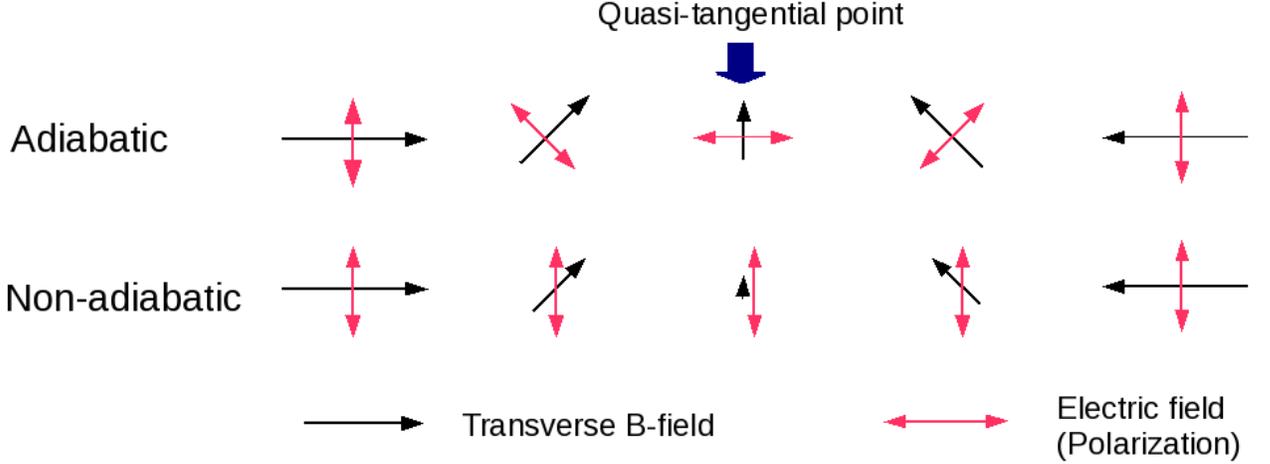,angle=0,height=7cm} \\
\end{tabular}
\caption{A sketch of the polarization mode evolution across the
  quasi-tangential point in the adiabatic (upper panels) and
  non-adiabatic (lower panels) limits. In both cases, the photon
  polarization vector $\vecE$ (double-arrowed bars) and the transverse
  magnetic component of the magnetic field $\vecB_\perp$
  (single-arrowed bars) are shown at five different positions along
  the ray (from left to right): before, slightly before, at, slightly
  after, after the QT point. In both limiting cases, there is no net
  change in the linear polarization when the photon traverses the QT
  region. Polarization change occurs only in the intermediate cases
  ($\Gamma_{\rm t}\sim1$; see text).
\label{fig:sketch}}
\end{figure*}

Before and after the QT point (where $s=0$), when $|s|\gg
\epsilon{\cal R}$, the adiabaticity parameter $\Gamma_{\rm ad}$
increases rapidly with $|s|$ and the mode evolution is generally
adiabatic ($\Gamma_{\rm ad}\gg1$).  However at the QT point,
$\Gamma_{\rm ad}$ reaches its minimum value:
\begin{equation}
\Gamma_{\rm t}\simeq
1.0\times10^8E_1B_{13}^2\epsilon^3{\cal R}_1.\label{eq:Gamma_t}
\end{equation}
The value of $\Gamma_{\rm t}$ determines the mode evolution
characteristics across the QT region.  Figure~\ref{fig:sketch} shows
the qualitative behaviors in two limiting cases: the adiabatic limit
($\Gamma_{\rm t}\gg 1$) and the non-adiabatic limit ($\Gamma_{\rm
  t}\ll 1$). In the adiabatic case, the photon polarization direction
follows the variation of the transverse magnetic field
$\vecB_\perp$. Since the final direction of $\vecB_\perp$ is opposite
to the initial direction, the final polarization direction is the same
as the initial one.  In the non-adiabatic case, $\varphi_B$ changes
rapidly, so that the polarization direction cannot follow
$\vecB_\perp$ and remains constant across the QT region. Thus, in both
limiting cases, the polarization direction is unchanged when the
photon traverses the QT point.

\begin{figure*}
\begin{tabular}{c}
\psfig{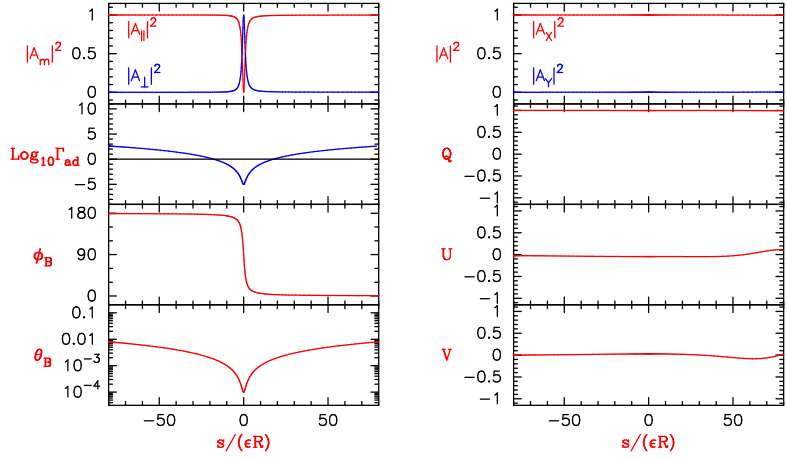} \\
\end{tabular}
\caption{Evolution of photon polarization across the quasi-tangential
  region for the $\Gamma_{\rm t}\ll 1$ case. The parameters $\epsilon,~{\cal
    R}$ are defined in equation~(\ref{eq:simplemode}), $s$ is the
  distance along the ray ($s=0$ corresponds to the QT point). Prior to
  the QT point, the photon is assumed to be in the $\parallel$-mode,
  with $A_\parallel=1$ and $A_\perp=0$. The angles
  $\theta_B,~\varphi_B$ specify the magnetic field orientation, $A_X$
  and $A_Y$ are the photon polarization amplitudes along the fixed
  $X,~Y$ axis, $Q,~U,~V$ are the Stokes parameters.  For this example,
  the parameters are $B=10^{12}$\,G, $E=1$\,keV, $\epsilon =
  5\times10^{-5}$, ${\cal R}=100$\,km.  Note that at the QT
  point, $\theta_B\sim0$, $\varphi_B$ varies rapidly, giving rise to
  non-adiabatic mode evolution.  }\label{fig:single_nonad}
\end{figure*}

\begin{figure*}
\begin{tabular}{c}
\psfig{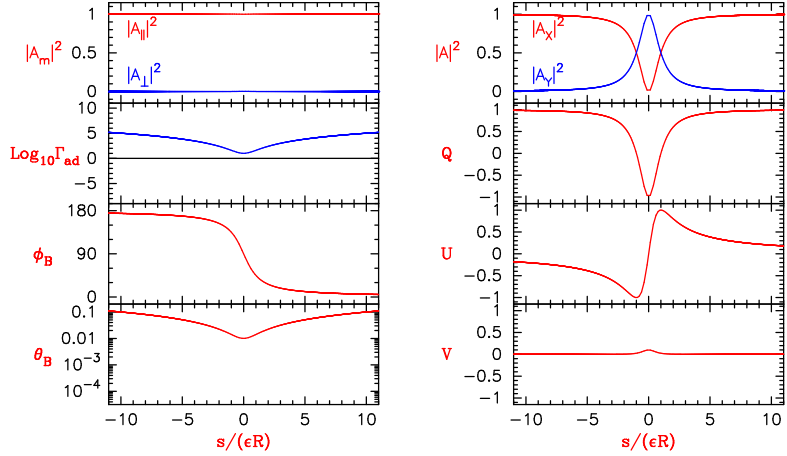} \\
\end{tabular}
\caption{Same as in Fig.~\ref{fig:single_nonad}, except for the
  $\Gamma_{\rm t}\gg 1$ case, with $\epsilon = 10^{-2}$. Note that in this
  case, at the QT point $\theta_B$ is not so close to $0$ and
  $\varphi_B$ varies slowly, giving rise to adiabatic mode evolution.
}\label{fig:single_ad}
\end{figure*}

\begin{figure*}
\begin{tabular}{c}
\psfig{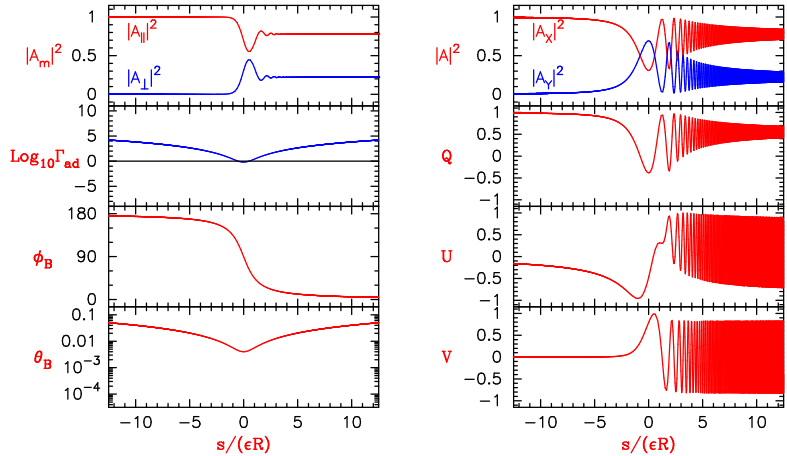} \\
\end{tabular}
\caption{Same as in Fig.~\ref{fig:single_nonad}, except for the
  $\Gamma_{\rm t}\sim 1$ case, with $\epsilon = 4\times10^{-3}$.  Note that
  in this case, partial mode conversion occurs at the QT point, and
  the oscillatory behaviors in the Stokes parameters are the result of
  interference between the two photon modes.
}\label{fig:single_middle}
\end{figure*}

Obviously, a non-trivial change of photon polarization across the QT
point occurs only in the intermediate case, $\Gamma_{\rm t}\sim 1$. To
obtain the quantitative behaviors of the polarization evolution for
general values of $\Gamma_{\rm t}$, we integrate eq.~(\ref{eq:me})
numerically.
Figures~\ref{fig:single_nonad}~--~\ref{fig:single_middle} show three
examples of single photon mode evolution in the QT region,
corresponding to $\Gamma_{\rm t}\ll 1$, $\Gamma_{\rm t}\gg 1$ and
$\Gamma_{\rm t}\sim 1$, respectively.  For the $\Gamma_{\rm t}\ll 1$
(non-adiabatic) case (Fig.~\ref{fig:single_nonad}), the photon
polarizations (Stokes parameters) are constant throughout the region.
For the $\Gamma_{\rm t}\gg 1$ (adiabatic) case
(Fig.~\ref{fig:single_ad}), the photon polarizations (Stokes
parameters) change around the QT region, following the variation of
$\vecB$, but the final polarizations are very close to the initial
values. The most interesting case occurs for $\Gamma_{\rm t}\sim 1$
(Fig.~\ref{fig:single_middle}). In this intermediate regime, partial
mode conversion takes place, so that after crossing the QT point, the
photon becomes a mixture of two modes (even when it is in a pure mode
prior to QT crossing). The polarization state of the photon is
therefore significantly changed by the QT effect.

Based on the above results, we can use $\Gamma_{\rm t}\lo 1$ to define
the parameter regime for which propagation through the QT region gives
rise to an appreciable change in the photon polarization,
\begin{equation}
\epsilon\lo \epsilon_{\rm crit}=2.15\times10^{-3}(E_1B_{13}^2{\cal R}_1)^{-1/3}.
\end{equation}
Note that when $\epsilon\ll \epsilon_{\rm crit}$, the QT effect is
also negligible. This equation effectively maps out the domain where
the QT propagation effect must be carefully considered. Outside this
domain, one can ignore the QT effect in determining the final observed
polarization signals.  To translate this effective parameter domain
into the physical size of the emission region requires a knowledge of
the global NS magnetic field structure.  The smallness of
$\epsilon_{\rm crit}$ for typical parameters (e.g., $E_1\sim
1,~B_{13}\sim 1$ and ${\cal R}_1\sim 1$) indicates that the
``affected'' region is a small fraction of the NS surface.  But as far
as the observed polarization signals are concerned, it is more
relevant to compare the size of the ``affected'' region to the size of
the photon emission area.  We consider the special case of dipole
magnetic field in the next section.

\section{Quasi-Tangential Effect in Dipole Magnetic Field}

In this section we assume that the NS has a pure dipole field. We
focus on X-ray emission from the polar cap region of the star
(sections 4.2-4.3), but also consider more general emission regions
on the NS surface (section 4.4).

\begin{figure*}
\begin{tabular}{cc}
\psfig{figure=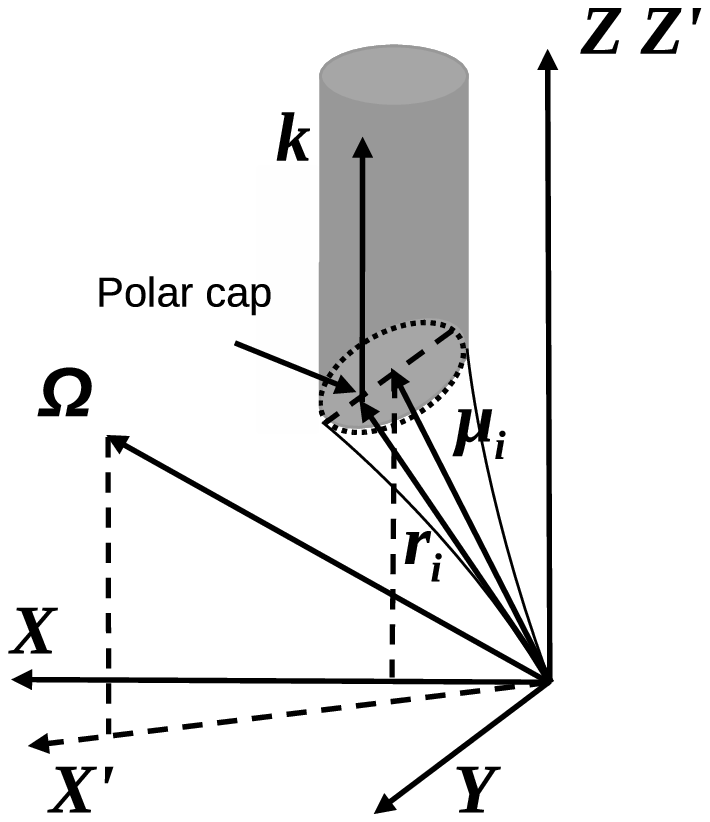,angle=0,height=8cm} &
\psfig{figure=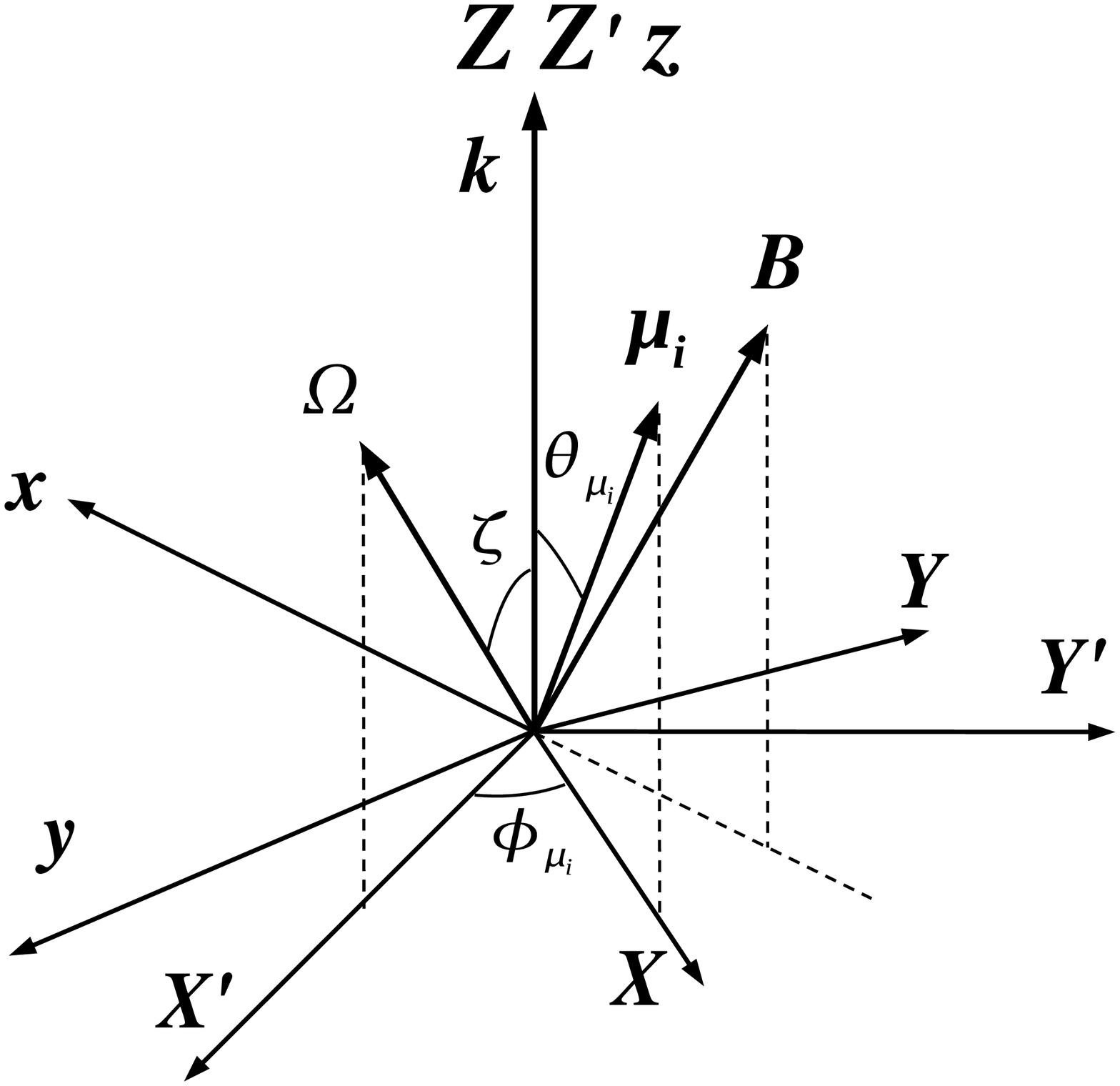,angle=0,height=8cm} \\
(a) & (b)\\
\end{tabular}
\caption{Geometrical model of X-ray emission in a dipole magnetic
  field. Photons are emitted from the polar cap region and propagate
  toward the observer in the direction $\veck$ along the $Z$ or
  $Z'$-axis. At the time of emission, the magnetic dipole vector is
  $\vecmui$ and the initial photon position is ${\bf r}_{\rm i}$.  Three
  coordinate systems are defined: (i) The fixed frame $XYZ$ with
  $\vechatZ\parallel\vechatk$, $\vecmui$ in the $XZ$-plane and
  $\vecmui=\mu(\sin\theta_\mui, 0, \cos\theta_\mui)$; (ii) the fixed
  frame $X'Y'Z'$ with $\vechatZ'\parallel\veck$, $\vechatW$ in the
  $X'Z'$-plane and $\vechatk\times\vechatW=\sin\zeta\,\vechatY'$;
  (iii) the instantaneous frame $xyz$, with
  $\vechatz\parallel\vechatk$, $\vecB$ in the $xz$-plane and
  $\vechatk\times\vechatB = -\sin\theta_B \vechaty$.  Note that as the
  photon propagates, the magnetic field it ``sees'' changes and thus
  the $xyz$ frame rotates around the $Z$-axis.
\label{fig:embeam}}
\end{figure*}

\subsection{Magnetosphere Field Geometry Along the Ray}

To calculate the observed polarized radiation signals, we set up a
fixed coordinate system $X'Y'Z'$ with the $Z'$-axis along the
line-of-sight (pointing from the NS toward the observer) and the
$X'$-axis in the plane spanned by the $Z'$-axis and $\vecW$ (the spin
angular velocity vector) (see Fig.~\ref{fig:embeam}). The angle
between $\vecW$ and $\veck$ is $\zeta$. The magnetic dipole $\vecmu$
rotates around $\vecW$, and $\alpha$ is the angle between $\vecW$ and
$\vecmu$.

Consider a photon emitted at time $t_{\rm i}$ (corresponding to the NS
rotation phase $\Psi_{\rm i}$) from the position $\vecr_{\rm i}=(R_\ast,
\theta_\ri, \varphi_\ri)$ on the NS surface.  At this emission time,
the magnetic dipole moment is $\vecmu_{\rm i}$.  As the photon propagates
along the $Z$-axis, its position vector changes as
\begin{equation}
\vecr = \vecr_{\rm i} + s \vechatZ, \label{eq:r}
\end{equation}
where $s=c(t-t_{\rm i})$ is the photon displacement from the emission point.
\footnote{Equation~\ref{eq:r} neglects the effect of gravitational
light bending, which can be incorporated in a straightforward manner
(e.g. Beloborodov 2002; van Adelsberg \& Lai 2006). This effect amounts
to shifting $\vecr_{\rm i}$ in the direction perpendicular to the $Z$-axis,
and does not appreciably change our result.}
In the meantime, $\vecmu$ rotates around $\vecW$, and changes
according to
\begin{equation}
\vecmu(s) = \mu\lb(\sin\zeta\cos\alpha+\cos\zeta\sin\alpha\cos\Psi)\vechatX' +
          \sin\alpha\sin\Psi \vechatY' +
      (\cos\zeta\cos\alpha - \sin\zeta\sin\alpha\cos\Psi)\vechatZ'\rb,
\end{equation}
where the rotation phase $\Psi$ is
(we set $\Psi=0$ when $\vecmu$ lies in the $X'Z'$ plane)
\begin{equation}
\Psi(s) = \Psi_{\rm i} + \Omega (t-t_{\rm i}) = \Psi_{\rm i} + s/r_{\rm lc},
\end{equation}
with $r_{\rm lc}=c/\Omega$ the radius of the light cylinder.
The polar angles $(\theta_\mu$, $\varphi_\mu)$ of $\vecmu$
in the $X'Y'Z'$ frame are given by
\begin{equation}
\cos\theta_\mu=\cos\zeta\cos\alpha - \sin\zeta\sin\alpha\cos\Psi,\quad
\tan\varphi_\mu=\frac{\sin\alpha\sin\Psi}{\sin\zeta\cos\alpha
+\cos\zeta\sin\alpha\cos\Psi}.
\label{eq:thetaphi_mu}
\end{equation}
(Similar expressions hold for the polar angles of $\vecmu_{\rm i}$,
$\theta_\mui$ and $\varphi_\mui$, with $\Psi$ replaced by $\Psi_{\rm
  i}$.)  The changing magnetic field as ``seen'' by the photon is
obtained from
\begin{equation}
\vecB(s)=-\nabla (\vecmu\cdot\vecr/r^3)
     =-\frac{\vecmu}{r^3}+\frac{3\vecr}{r^5}(\vecmu\cdot\vecr). \label{eq:B}
\end{equation}

When discussing the polarization result of a given rotation phase (at
emission) $\Psi_{\rm i}$, it is convenient to introduce another fixed
coordinate system $XYZ$ (see Fig.~\ref{fig:embeam}), such that
$\vechatZ\parallel\vechatk$ and $\vecmui$ in the $XZ$-plane with
$\vecmui=\mu(\sin\theta_\mui, 0, \cos\theta_\mui)$. In the $XYZ$
frame, $\varphi_\mui$ is given by
\begin{equation}
\vecmu(s) = \mu\lb \sin\theta_\mu\cos(\varphi_\mu-\varphi_\mui)\vechatX+
            \sin\theta_\mu\sin(\varphi_\mu-\varphi_\mui)\vechatY+\cos\theta_\mu\vechatZ\rb.
\end{equation}
The magnetic field (\ref{eq:B}) as ``seen'' by the photon is inclined
at an angle $\theta_B$ with respect to the line-of-sight, and makes an
azimuthal angle $\varphi_B$ in the $XY$-plane. The angle $\theta_B$,
$\varphi_B$ can be obtained from equation~\ref{eq:B} via:
\begin{equation}
\cos\theta_B(s) = \frac{B_Z}{B}, \quad
\tan\varphi_B(s) = \frac{B_Y}{B_X}. \label{eq:theta_B}
\end{equation}

\subsection{Polarization Map of the Polar Cap Emission}
\label{sec:sim}

The observed polarized radiation is the incoherent sum of photons from
the emission region on the NS surface. For each emission point,
$\vecr_{\rm i}=(R_\ast,~\theta_\ri,~\varphi_\ri)$, we integrate the mode
evolution equation (\ref{eq:me}) along the photon ray from $\vecr_{\rm i}$
to a large radius, beyond the {\it polarization limiting radius}, to
determine the final polarization state of the photon. The polarization
limiting radius, $r_{\rm pl}$, is where the two photon modes start
recoupling to each other, and is determined by the condition
$(\omega/c)\Delta n=2|d\varphi_B/ds|$.  At large distance ($r\gg R_\ast$),
the magnetic field is simply $\vecB\simeq \left(2\mu_Z{\hat
  Z}-\mu_X{\hat X}-\mu_Y{\hat Y}\right)/r^3$, and $r_{\rm pl}$ is
given by
\begin{equation}
{r_{\rm pl}\over R_\ast}\simeq 70 \left( E_1 B_{\ast 13}^2P_1\right)^{1/6},
\end{equation}
where $B_{\ast 13}$ is the polar magnetic field at the stellar surface
in units of $10^{13}$\,G (see van Adelsberg \& Lai 2006 for a more
detailed expression).  Beyond $r_{\rm pl}$, the photon polarization
state is frozen.  As mentioned in section 1, the calculations
presented in van Adelsberg \& Lai (2006) and Lai \& Ho (2003b) did not
consider the possibility that the photon polarization may change
appreciably when crossing the QT point, which typically lies at a much
smaller radius than $r_{\rm pl}$.

In general, the radiation emerging from the NS atmosphere at $\vecr_{\rm i}$
includes both the $\parallel$-mode and the $\perp$-mode, with the
intensities $I_\parallel,~I_\perp$ depending on the field strength,
photon energy and emission angle (see Lai \& Ho 2003b and van
Adelsberg \& Lai 2006). In the absence of the QT effect, the radiation
at $r>r_{\rm pl}$ will consist of approximately the same $I_\parallel$
and $I_\perp$, with a small mixture of circular polarization generated
around $r_{\rm pl}$ (see van Adelsberg \& Lai 2006). This simple
result should be modified if there is a significant polarization
change when the photon crosses the QT region. Since we are interested
in understanding the QT effect, in the following we will assume that
at the emission point the radiation is in the $\parallel$-mode, with
$I_\parallel=1$ and $I_\perp=0$.

\begin{figure*}
\begin{tabular}{c}
\psfig{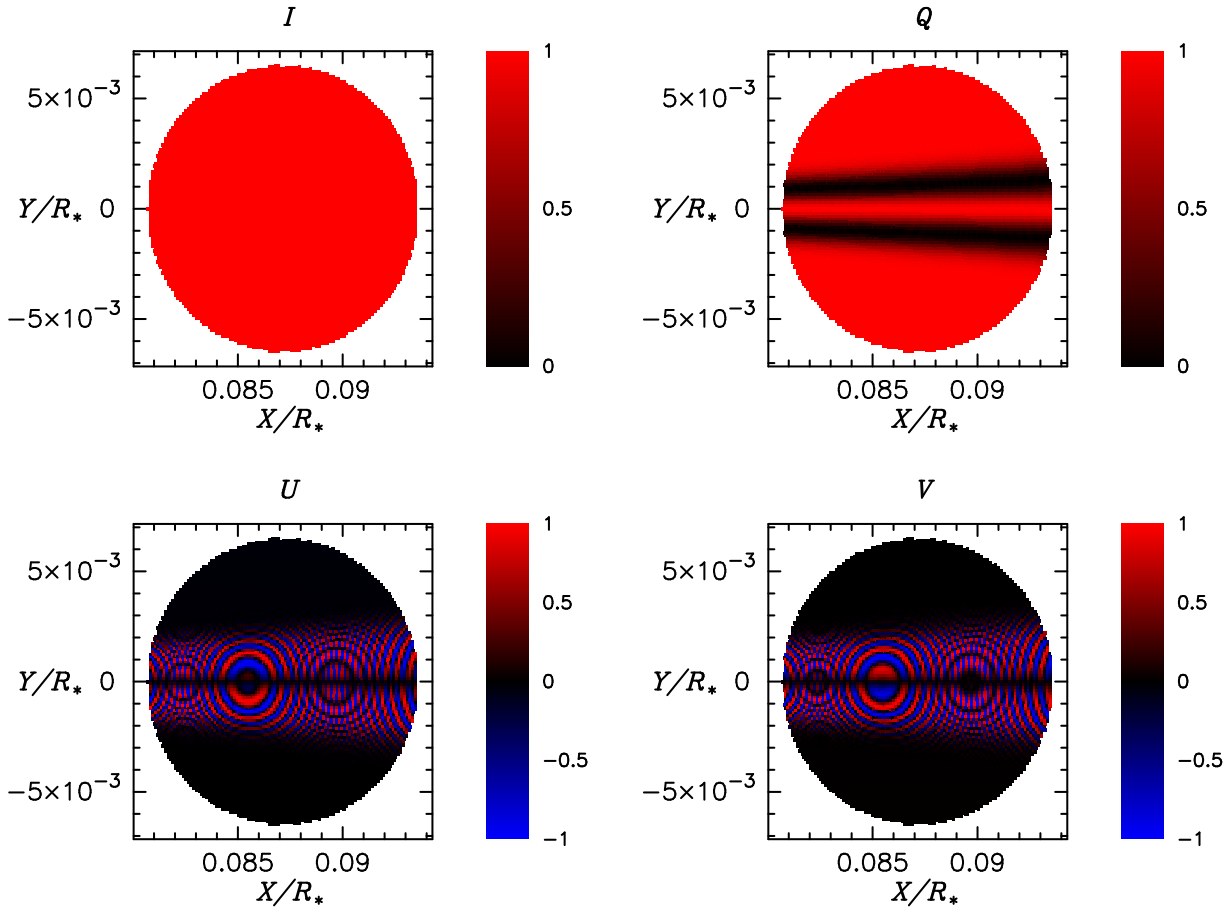} \\
\end{tabular}
\caption{Two-dimensional polarization map of emission from the polar
  cap region of a NS. The QT propagation effect changes the photon
  polarization in two narrow bands (confirmed to $-W_{\rm
    t}/2<Y<W_{\rm t}/2$; see Fig.~\ref{fig:2D_3t}). The input
  parameters are: $B_\ast=10^{14}\,$G, $E=1\,$keV and
  $\theta_\mui=5^o$ (We choose $P=5\,$s, $\alpha=30^o$, $\zeta=35^o$
  and $\Psi_{\rm i}=180^o$, although the map depends very weakly on
  these parameters as long as $\theta_\mui$ is the same).
\label{fig:2D}}
\end{figure*}

\begin{figure*}
\begin{tabular}{c}
\psfig{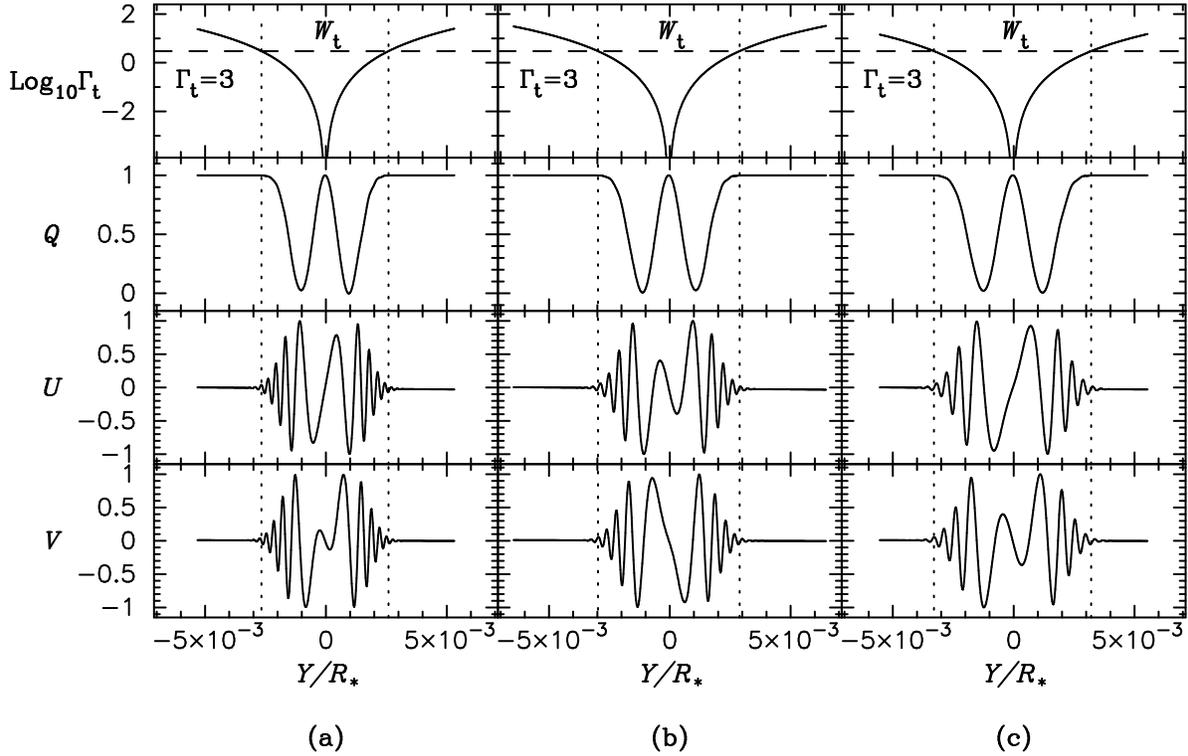} \\
\end{tabular}
\caption{One-dimensional polarization profiles for emission from the
  polar cap region of a NS. These profiles are produced by fixing
  $\theta_\ri$ while varying $\varphi_\ri$ of the emission point. The
  coordinates $XY$ are the projection of the emission point
  $\vecr_{\rm i}$ on the $XY$ plane (see Fig.~5). The three columns
  (from left to right) correspond to a fixed $\theta_\ri=4.8^o$,
  $5.0^o$, $5.2^o$ (or $X/R_\ast\simeq 0.837$, 0.872, 0.906),
  respectively.  In each column, the final Stokes parameters $Q,~U,~V$
  are plotted as a function of $Y$, as well as $\Gamma_{\rm t}$, the
  adiabaticity parameter at the QT point. The polarizations of photons
  emitted from the region $-W_{\rm t}/2<Y<W_{\rm t}/2$ (between the
  two vertical dashed lines, defined by $\Gamma_{\rm t}=3$) are
  modified by the QT effect. The input parameters for these profiles
  are: $B_\ast=10^{14}\,$G, $E=1\,$keV and $\theta_\mui=5.0^o$ (We
  choose $P=5\,$s, $\alpha=30^o$, $\zeta=35^o$, $\Psi_{\rm i}=180^o$,
  although the profiles depend very weakly on these parameters as
  long as $\theta_\mui$ is the same).
 \label{fig:2D_3t}}
\end{figure*}

\begin{figure*}
\begin{tabular}{c}
\psfig{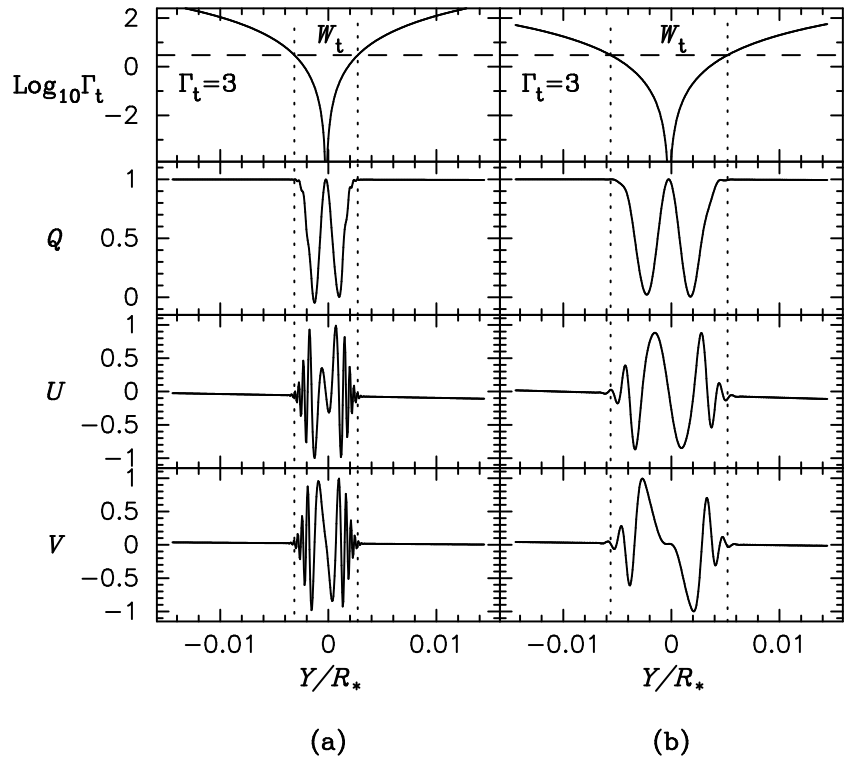} \\
\end{tabular}
\caption{Same as Fig.~\ref{fig:2D_3t}, except for $B_\ast=10^{14}$\,G
  (the left panels) and $B_\ast=4\times10^{13}$\,G (the right panels),
  both for $P=1$\,s.
\label{fig:1D_B}}
\end{figure*}

Figure~\ref{fig:2D} gives an example of the 2-dimensional polarization
map of the final Stokes parameters of photons emitted from the NS
polar cap region. This map is produced for a specific set of
parameters: surface magnetic field $B_\ast$, photon energy $E$, and
the angle between $\vecmui$ and $\veck$, $\theta_\mui$. [For a
  rotating NS, the map also depends on the angles $\alpha$, $\zeta$,
  the rotation period $P$ and emission phase $\Psi_{\rm i}$, but the
  dependence is mainly through $\theta_\mui$ --- see
  eq.~(\ref{eq:thetaphi_mu}). see below.]  Without the QT effect, the
final polarization from the polar cap region will be $Q=1$.  We see
from Fig.~\ref{fig:2D} that the QT effect gives rise to fine features
(with length scale much less than the polar cap size) in the
polarization map.  In particular, there are two narrow bands (which
are parallel to the projected magnetic dipole axis) in the emission
region, where the linear polarization $Q$ is significantly reduced
because the original $\parallel$ mode is converted across the QT
region into a mixture of $\parallel$-mode and $\perp$-mode. In these
two narrow belts, non-zero pattens of Stokes $U$ and $V$ are
produced. However, these $U,~V$ patterns are anti-symmetric with
respect to the $Y=0$ line (i.e., the $\veck$-$\vecmui$ plane), so that
the observed values of $U$ and $V$ almost equal zero when photons from
the whole emission region are added.

In Fig.~\ref{fig:2D_3t} we show three sections of Fig.~\ref{fig:2D}
along the $Y$-axis for a fixed value of $X$ [$X/R_\ast=0.0837$ or
  $\theta_\ri=4.8^o$ for panel (a), $X/R_\ast=0.0872$ or
  $\theta_\ri=\theta_\mui=5^o$ for panel (b), and $X/R_\ast=0.0906$ or
  $\theta_\ri=5.2^o$ for panel (c)].  We also plot the adiabaticity
parameter $\Gamma_{\rm t}$ at the QT point for the photon rays from
each emission point. The $\Gamma_{\rm t}$ profile is useful for
understanding the results for the Stokes parameters.  As the analysis
in section \ref{sec:qte} shows, photon propagation through the QT
region changes the photon polarization significantly only when
$\Gamma_{\rm t}\sim 1$. From Fig.~\ref{fig:2D_3t}, we see that the
photon Stokes parameters are modified by the QT effect only in two
narrow bands, confined in the region $-W_{\rm t}/2<Y<W_{\rm t}/2$,
with the boundaries defined approximately by $\Gamma_{\rm t}=3$ (the
two vertical dotted lines). Outside this region, $\Gamma_{\rm t}\gg
1$, the mode evolution is adiabatic around the QT point, so that the
photon polarization state follows the magnetic field direction and the
Stokes parameters are unchanged across the QT region.  In the middle
of the band (i.e., around $Y=0$), $\Gamma_{\rm t}\ll1$, mode evolution
is non-adiabatic, and the photon polarization state also doesn't
change across the QT point (see Fig.~\ref{fig:sketch}).

Figure~\ref{fig:1D_B} depicts other examples of the 1D polarization
map for different values of stellar magnetic field $B_\ast$ ($10^{14}$
and $4\times 10^{13}$\,G) and spin period $P$. As in
Fig.~\ref{fig:2D_3t}, we see that the final photon polarizations are
determined by the value of $\Gamma_{\rm t}$, the QT effective region
is confined to $-W_{\rm t}/2<Y<W_{\rm t}/2$ defined by $\Gamma_{\rm t}<3$ (the
region between the two vertical dotted lines).

\begin{figure*}
\begin{tabular}{c}
\psfig{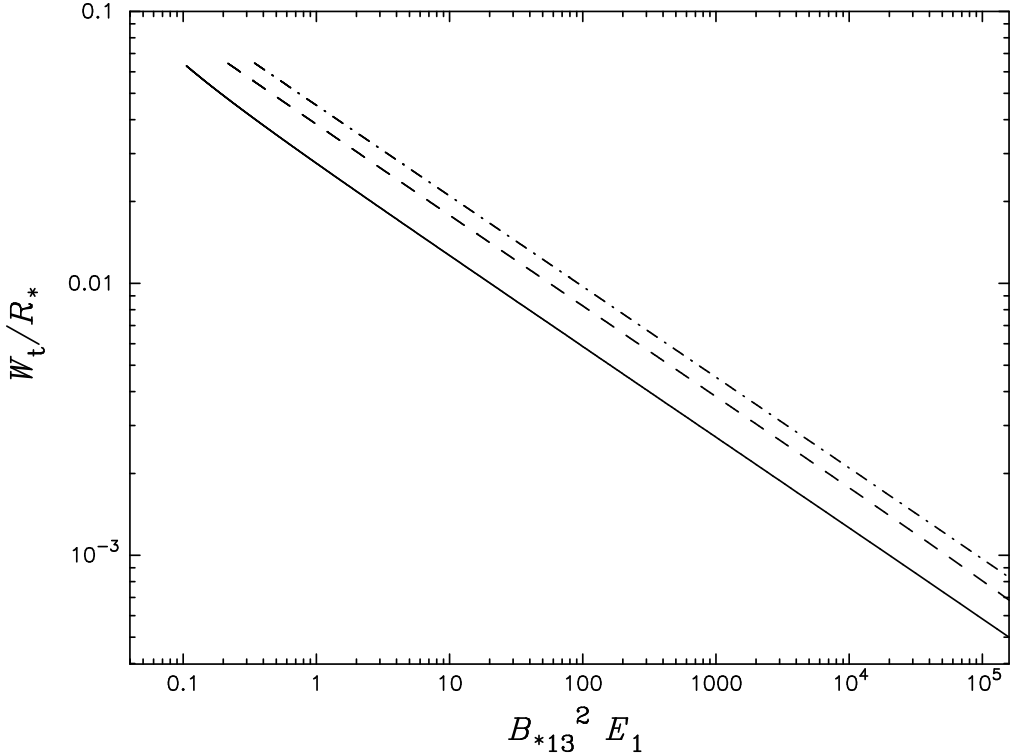} \\
\end{tabular}
\caption{The width of the QT effective region $-W_{\rm t}/2<Y<W_{\rm t}/2$,
  defined by $\Gamma_{\rm t}<3$, as a function of the surface magnetic field
  strength and photon energy.  Here $B_{\ast14}=B_\ast/(10^{14}\,\rm
  G)$, $E_1=E/(1\,\rm keV)$. The different lines correspond to
  different $\theta_\mui$: $\theta_\mui=5^o$ (solid line),
  $\theta_\mui=15^o$ (dashed line), $\theta_\mui=45^o$
  (dot-dashed line).
\label{fig:wd_B}}
\end{figure*}

A careful examination of Figs.~\ref{fig:2D_3t}~--~\ref{fig:1D_B} shows
that due to the NS rotation, the $\Gamma_{\rm t}$ profile and Stokes
profiles shift in $Y$ by the amount $\Delta Y \simeq
(R_\ast\sin\alpha/r_{\rm lc})s_{\rm t}$, where $s_{\rm t}$ is the
distance of the QT point from the emission point. Since typically
$s_{\rm t}$ is less than a few NS radii, this shift $\Delta Y$ is
negligible.  According to eq.~(\ref{eq:Gamma_t}), $\Gamma_{\rm t}$ is
proportional to $EB_\ast^2$, thus for smaller $B_\ast$ (and smaller
$E$) the effective width $W_{\rm t}$ is larger, as seen in
Fig.~\ref{fig:1D_B}.  In Fig.~\ref{fig:wd_B} we show how $W_{\rm t}$
changes with varying $EB_\ast^2$. Our numerical result for $W_{\rm
  t}$ can be fitted by
\begin{equation}
\frac{W_{\rm t}}{R_\ast} \simeq
2.7\times10^{-2}\,(B_{\ast13}^2E_{1})^{-1/3}f(\theta_\mui).
\label{eq:W}
\end{equation}
Here $f(\theta_\mui)$ is a dimensionless function of $\theta_\mui$
[which in turn depends on $\alpha$, $\zeta$ and $\Psi_{\rm i}$, see
  eq.(\ref{eq:thetaphi_mu})]: $f(\theta_\mui=5^o)= 1$ and varies from
1 to 1.7 for different values of $\theta_\mui$ (see
Fig.~\ref{fig:f_thetamui}).  The scaling relation in eq.~(\ref{eq:W})
is derived in the appendix.

Note that the polar cap width (diameter) is given by
\begin{equation}
{W_{\rm cap}\over R_\ast}=2\sqrt{\frac{R_\ast}{c/\Omega}}\simeq
1.3\times10^{-2}P_5^{-1/2}, \label{eq:polarcap}
\end{equation}
where $P_5=P/(5{\rm s})$. Then we have
\begin{equation}
\frac{W_{\rm t}}{W_{\rm cap}} =
2.1(B_{\ast13}^2E_{1}P_5^{-3/2})^{-1/3}f(\theta_\mui).
\label{eq:W/Wcap}
\end{equation}
Equation~(\ref{eq:W/Wcap}) implies that the size of the effective QT
region (where the QT effect changes the photon polarization) can be
comparable to the polar cap size for some parameters (e.g., low photon
energy and low magnetic field strength).

\begin{figure*}
\begin{tabular}{c}
\psfig{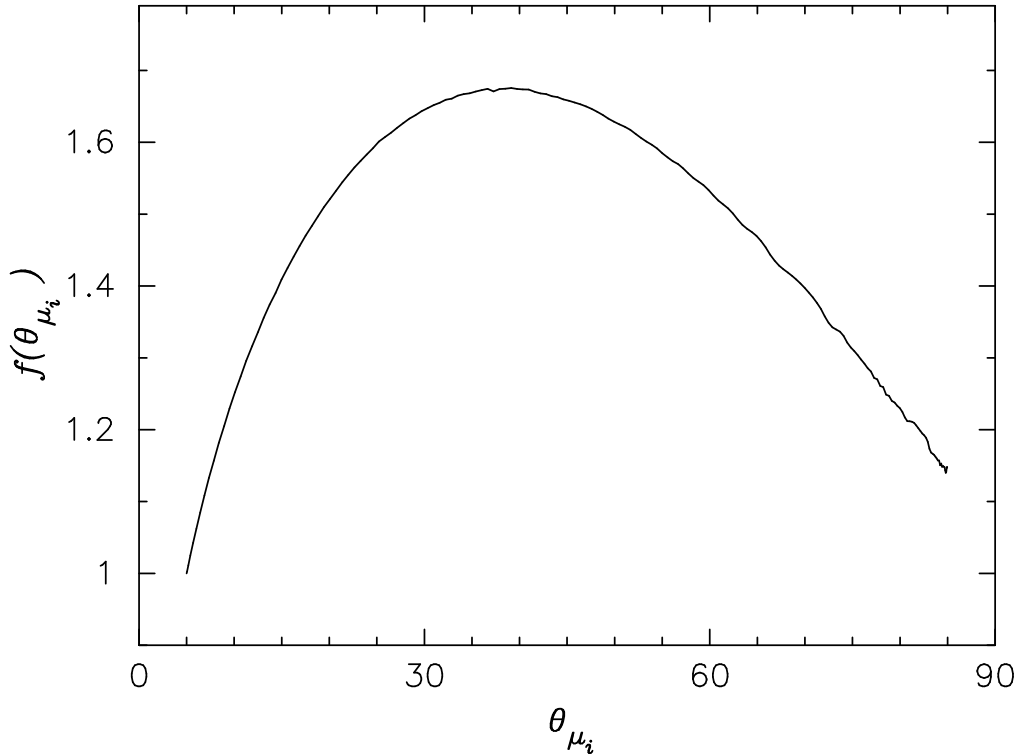} \\
\end{tabular}
\caption{The dimensionless function $f(\theta_\mui)$,
as defined in eq.~(\ref{eq:W}). 
\label{fig:f_thetamui}}
\end{figure*}

\subsection{Quasi-Tangential Effect on the Observed Polarization}

\begin{figure*}
\begin{tabular}{c}
\psfig{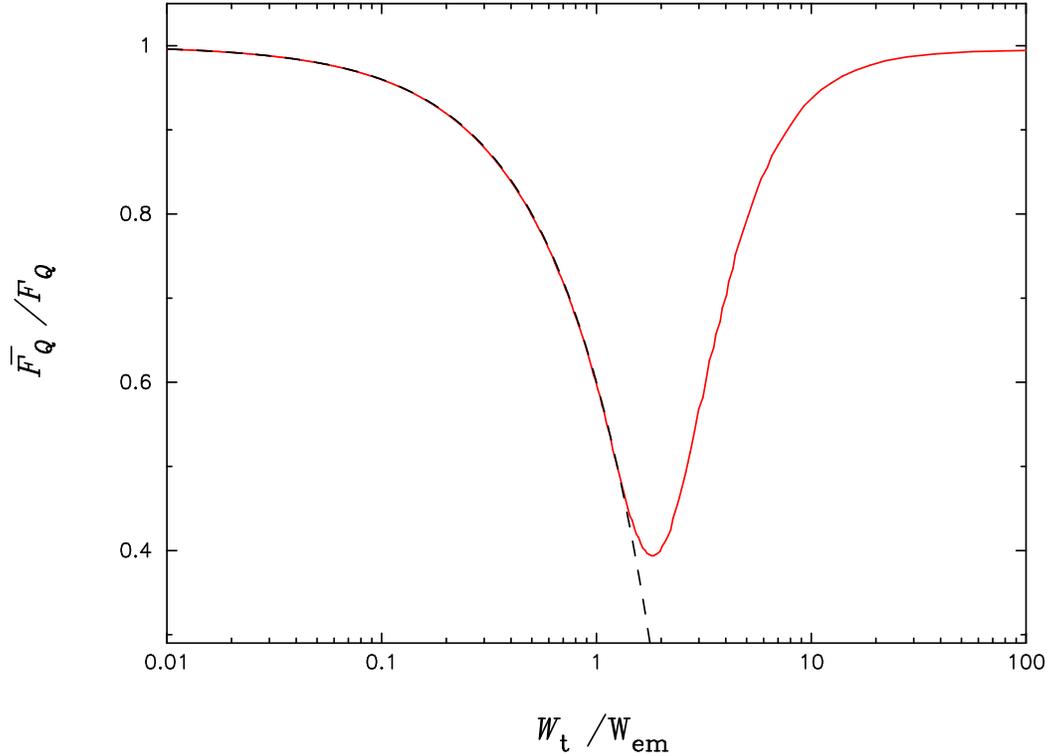} \\
\end{tabular}
\caption{The effect of QT propagation on the observed linear
  polarization: $F_Q$ ($\FQb $) is the linearly polarized radiation
  flux before (after) traversing the QT region; see
  eq.~(\ref{eq:FQ1}). Here $W_{\rm t}$ is the width of the effective
  QT region (where $\Gamma_{\rm t}<3$) and $W_{\rm em}$ is the width
  of the emission region. For $W_{\rm t}/W_{\rm em}\lo 1$, $\FQb /F_Q$
  can be approximated by eq.~(\ref{eq:FQ}), which is shown as the
  dashed line.
\label{fig:FL_W}}
\end{figure*}

As discussed above, when linearly polarized radiation with $Q=I=1$
traverses the QT region with $\Gamma_{\rm t}<3$, its Stokes parameters will
be changed (so that $Q$ will become less than unity, and $U,~V$ will
be nonzero; see Figs.~\ref{fig:2D_3t}~--~\ref{fig:1D_B}). However, when
adding up radiation from a finite emission region, we find $\int dY
U\simeq 0$ and $\int dY V\simeq 0$.  Thus the net effect of the QT
propagation is to reduce the degree of the linear polarization of the
photon. In general, if $F_Q$ is the flux of linearly polarized
radiation prior to passing the QT region, then after traversing the QT
region, the linearly polarized radiation flux $\FQb $ can be obtained
by
\begin{equation}
{\FQb \over F_Q}=\int_{|Y|<W_{\rm em}/2}\frac{Q(Y)}{I}\rd Y,\label{eq:FQ1}
\end{equation}
where $W_{\rm em}$ is the width of the emission region.  As seen in
section \ref{sec:sim} (see Figs.~\ref{fig:2D_3t}~--~\ref{fig:1D_B}), the
$Q(Y)$ profiles for different parameters are similar, so that $\FQb
/F_Q$ depends only on $W_{\rm t}/W_{\rm em}$.  Our numerical result for
$\FQb /F_Q$ as a function of $W_{\rm t}/W_{\rm em}$ is shown in
Fig.~\ref{fig:FL_W}. For a given set of parameters ($B_\ast$, $E$,
$\theta_\mui$), we can use eq.~(\ref{eq:W}) and Fig.~\ref{fig:f_thetamui} to
obtain $W_{\rm t}$, and then use Fig.~\ref{fig:FL_W} to read off the ratio 
$\FQb /F_Q$. If $W_{\rm t}\lo W_{\rm em}$, we find that $\FQb /F_Q$ is
approximately given by
\begin{equation}
\FQb /F_Q\simeq 1-\frac{W_{\rm t}}{2.5W_{\rm em}}\qquad
({\rm for}~~W_{\rm t}/W_{\rm em} \lo 1). \label{eq:FQ}
\end{equation}
Note that for $W_{\rm t}\ll W_{\rm em}$, the effective QT region is much
smaller compared to the emission region so that $\FQb /F_Q\simeq 1$; for
$W_{\rm em}\ll W_{\rm t}$, the photon mode evolution across the QT region is
non-adiabatic so that $\FQb /F_Q$ is also close to unity. Thus in both
$W_{\rm t}/W_{\rm}\ll 1$ and $W_{\rm t}/W_{\rm em}\gg 1$ limits, $\FQb /F_Q=1$,
i.e., the linear polarization is unchanged by the QT effect. The
minimum value of $\FQb /F_Q$ occurs at $W_{\rm t}/W_{\rm em}\simeq 1.8$.

\begin{figure*}
\begin{tabular}{c}
\psfig{figure=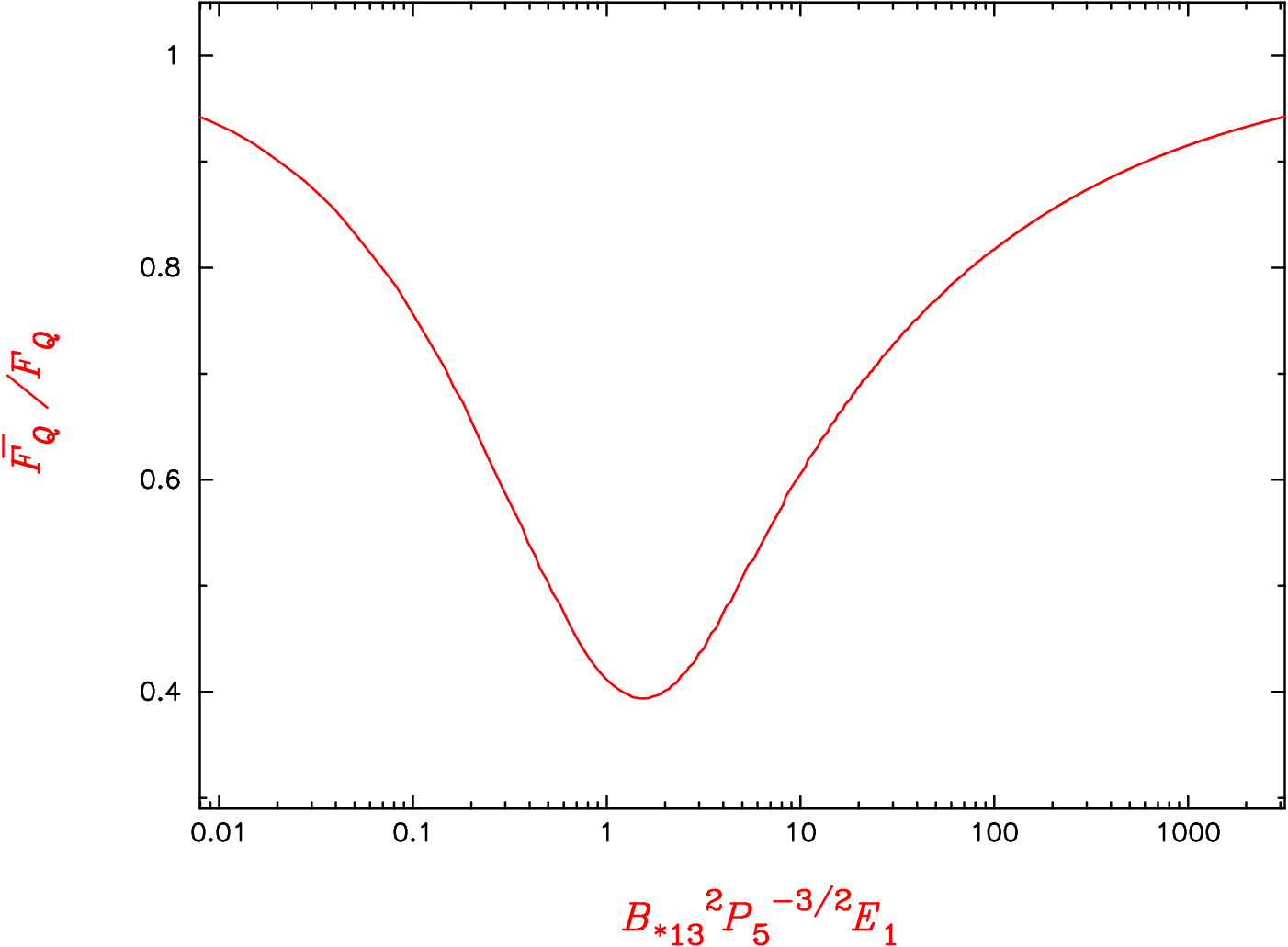,angle=-90,height=10cm} \\
\end{tabular}
\caption{The reduction factor $\FQb /F_Q$ of the linear polarization as
  a function of the photon energy (and other parameters).  Here
  $B_{\ast13}=B_\ast/(10^{13}\,\rm G)$, $E_1=E/(1\rm\,keV)$, $P_5=P/(5\,
  \rm s)$, and we have assumed $W_{\rm em}=W_{\rm cap}$.  The angle
  between $\veck$ and $\vecmui$ is chosen to be $\theta_{\mui}=5^o$.
\label{fig:spec}}
\end{figure*}

Suppose the emission is coming from polar cap region with width $W_{\rm
  em}=W_{\rm cap}$, then according to eq.~(\ref{eq:W/Wcap}), $W_{\rm
  t} /W_{\rm cap}$ is proportional to $(B_{\ast13}^2E_1)^{-1/3}P_5^{1/2}$.
Thus, for given $B_{\ast13}$, $P_5$ and $\theta_\mui$, we can
translate Fig.~\ref{fig:FL_W} into a spectrum for $\FQb /F_Q$ --- This
spectrum is shown in Fig.~\ref{fig:spec}.
We see that at sufficiently low energies and high energies,
$\FQb /F_Q\simeq 1$, and the minimum reduction of the degree of linear
polarization due to the QT effect occurs at $W_{\rm t}/W_{\rm cap}\simeq 1.8$,
corresponding to the photon energy
\begin{equation}
E_{\rm min}\simeq 1.6 B_{\ast13}^{-2}P_5^{3/2}f^3(\theta_\mui)\,{\rm keV}.
\label{eq:Emin}
\end{equation}

\begin{figure*}
\begin{tabular}{c}
\psfig{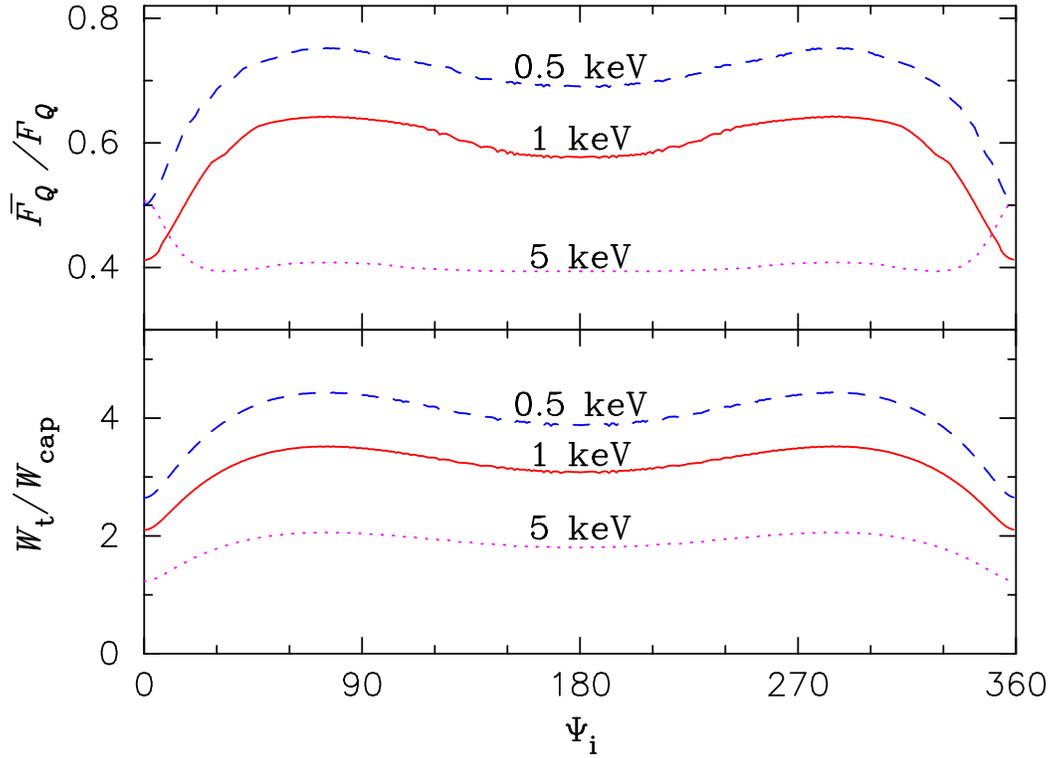} \\
\end{tabular}
\caption{The phase evolution of the modification of linear
  polarization by the QT effect, $F_Q'/F_Q$ (upper panel)
  and the effective QT region width relative to the polar cap size,
  $W/W_{\rm cap}$ (lower panel). The different curves are for
  different photon energies: $E=$0.5\,keV (dashed lines), 1\,keV (solid
  lines) and 5\,keV (dotted line).  The input model parameters are:
  $B_{\ast}=10^{13}$\,G, $P=5$\,s, $\alpha=30^o$ and $\zeta=35^o$. The
  emission region size is assumed to be $W_{\rm cap}$.
\label{fig:profile1}}
\end{figure*}

\begin{figure*}
\begin{tabular}{c}
\psfig{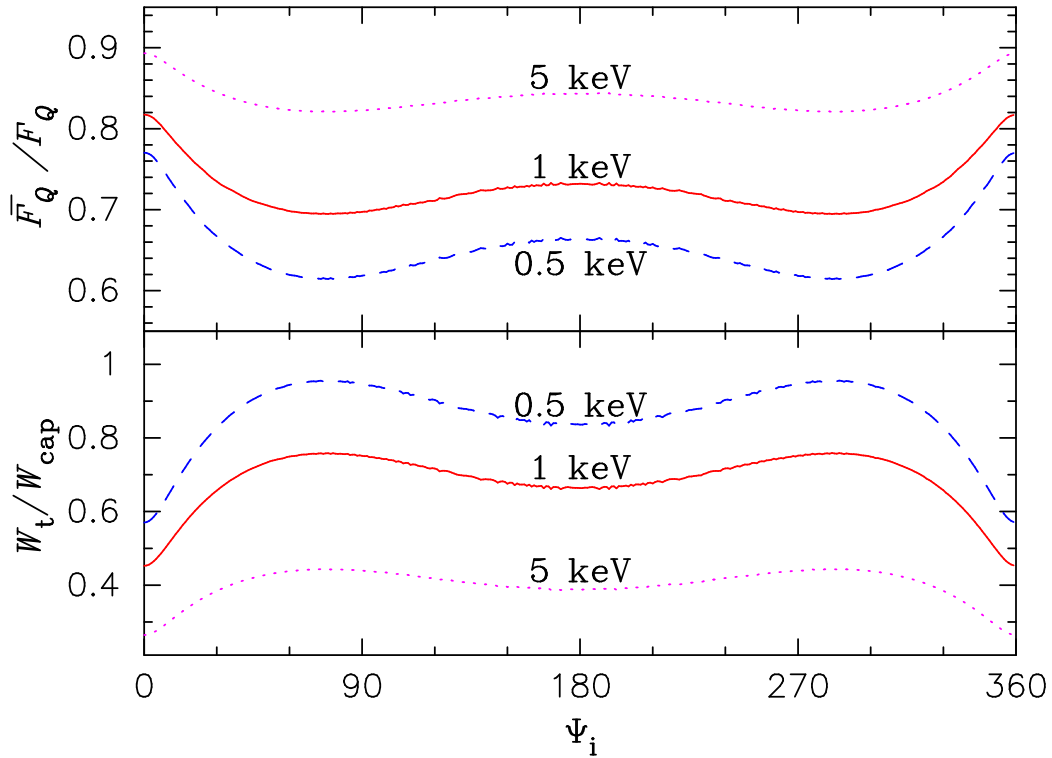} \\
\end{tabular}
\caption{Same as Fig.~\ref{fig:profile1}, except for
  $B_{\ast}=10^{14}$\,G.
\label{fig:profile2}}
\end{figure*}

Now we consider the (energy-dependent) polarization light curve
produced by the polar cap emission of a rotating NS. As the NS
rotates, the $\veck$-$\vecmui$ angle $\theta_\mui$ changes with the
rotation phase $\Psi_{\rm i}$ [see eq.~(\ref{eq:thetaphi_mu})].
According to eq.~(\ref{eq:W/Wcap}) and Figs.~\ref{fig:f_thetamui},
\ref{fig:FL_W}, different $\theta_\mui$ will give different $\FQb
/F_Q$ , which means $\FQb /F_Q$ will evolve with the rotation phase
$\Psi_{\rm i}$. Moreover, $W_{\rm t}/W_{\rm cap}$ and $\FQb /F_Q$
depend on the photon energy.  Figs.~\ref{fig:profile1} and
\ref{fig:profile2} present two examples of the phase evolution of
$W_{\rm t}/W_{\rm cap}$ and $\FQb /F_Q$, for $B_\ast= 10^{13}$\,G and
$10^{14}$\,G, respectively.  As noted before, $\FQb /F_Q$ reaches a
minimum at $W_{\rm t}/W_{\rm cap}= 1.8$ (see Fig.~\ref{fig:FL_W}).  In
the case depicted in Fig.~\ref{fig:profile1}, $W_{\rm t}/W_{\rm
  cap}>1.8$ for $E=1$\,keV and 0.5\,keV at all phases, thus $\FQb
/F_Q$ is larger for lower photon energies (since $W_{\rm t}/W_{\rm
  cap}$ increases with $E$).  In Fig.~\ref{fig:profile2}, $W_{\rm
  t}/W_{\rm cap}<1.8$ for all the three photon energies ($E
=0.5,~1,~5$\,keV), so that $\FQb /F_Q$ is larger for higher energies.


To produce the observed polarized radiation fluxes, using the results
present above, we will need input $F_Q$. In general, the emergent
radiation from the NS atmosphere (before passing through the QT
region) is linearly polarized, with $F_Q$ dependent on $E$ and the
rotation phase (and $B_\ast$ $\alpha$, $\zeta$). In the $XYZ$ frame,
the radiation has polarized fluxes $F_Q\neq 0$ (which can be either
positive or negative), $F_U=0$ and $F_V=0$.  Using $F_Q$ from an
atmosphere model and our result for $\FQb /F_Q$, we can compute ${\bar
  F}_Q$, the polarized flux after passing through the QT region.  Then
the {\it observed} polarized radiation fluxes in the fixed $X'Y'Z'$
frame are given by
\begin{equation}
F_Q'={\bar F}_Q\cos 2\varphi_B(r_{\rm pl}),\quad
F_U'={\bar F}_Q\sin 2\varphi_B(r_{\rm pl}),
\end{equation}
where $\varphi_B(r_{\rm pl})$ is the azimuthal angle of the magnetic
field, $\varphi_B$, evaluated at the polarization limiting radius
$r=r_{\rm pl}$.  For $r_{\rm pl}\ll r_{\rm lc}=c/\Omega$, or for spin
frequency $\ll 70$~Hz, $\varphi_B(r_{\rm pl})\simeq \pi+\varphi_\mu
\simeq \pi+\varphi_\mui$ (see Lai \& Ho 2003b; van Adelsberg \& Lai
2006), thus $F_Q'=\FQb\cos2\varphi_\mui$ and
$F_U'=\FQb\sin2\varphi_\mui$.

\subsection{Quasi-Tangential Effect for Different Emission Regions}

In the previous subsections we have focused on emission from the polar
cap region of the NS. In reality, the ``hot spot'' on the NS surface may
be larger. The QT effect also exists outside the polar cap region.

As discussed above, the QT effective region is confined to
$\Gamma_{\rm t}<3$. For a given dipole field geometry, we can
calculate $\Gamma_{\rm t}$ for photon rays emerging from different
points on the stellar surface.  Figs.~\ref{fig:2D_theta_mu15} and
\ref{fig:2D_theta_mu5} give two examples of the effective QT region as
defined by $\Gamma_{\rm t}<3$, for the $\veck$-$\vecmui$ angle
$\theta_\mui=15^o$ and $5^o$, respectively.  For each emission point,
if we ignore the QT effect, the final photon polarization is dominated
by the wave mode coupling effect and the polarization position angle
(PA) is approximately determined by $\varphi_B$ at polarization
limiting radius $r_{\rm pl}$ (again, assuming at emission the photon
is in the $\parallel$-mode).  Generally we have $R_\ast\ll r_{\rm
  pl}\ll r_{\rm lc}$, so $\varphi_B(r_{\rm pl})\simeq \varphi_\mui
+\pi$.  Therefore, the final polarization direction of the X-ray
photon from any point of the star surface are always parallel to the
$\veck$-$\vecmu$ plane. However, in the QT effective region, the final
PA may be modified when the photon propagates through QT region (see
the lower panel of Fig.~\ref{fig:2D_theta_mu15}). Note that the final
polarization angles are always symmetric with respect to the
$\veck$-$\vecmu$ plane. Thus, as discussed before, when adding up
radiation from a finite emission region, the net effect of QT
propagation is to reduce the degree of linear polarization without
changing the polarization angle.

Special caution must be taken when $\theta_\mui$ is small.  We see
from Fig.~\ref{fig:2D_theta_mu5} that for $\theta_\mui=5^o$, the
effective QT region may cover a significant part of the stellar
surface (In Fig.~\ref{fig:2D_theta_mu5}, the hatched region only shows
the effective QT region that satisfies $\theta_\ri<20^o$.)  This can
be understood as follows.  For $R_\ast\ll r\ll r_{\rm lc}$ the
transverse parts of the dipole magnetic field are [see
  eq.~(\ref{eq:B})]
\begin{equation}
B_X=\frac{\mu}{r^3}\lp-\sin\theta_\mui+3X_{\rm i}\cos\theta_\mui/r\rp, \quad
B_Y=\frac{\mu}{r^3}\lp\frac{3Y_{\rm i}\cos\theta_\mui}{r}\rp. \label{eq:B_perp}
\end{equation}
with $\vecr=\vecr_{\rm i}+s\vechatZ$ (neglecting light bending) and
$s\gg r_{\rm i}$, and $X_{\rm i},~Y_{\rm i}$ are the $X,Y$ components
of the emission position $\vecr_{\rm i}$.  For $\theta_\mui\gg
3R_\ast/r$, we have $B_X\simeq -(\mu/r^3)\sin\theta_\mui$, $B_Y\simeq
0$, thus outside the QT region, the final polarization is aligned with
the $\veck-\vecmui$ plane. However, for $\theta_\mui\lo 3R_\ast/r$, we
no longer can neglect $B_Y$ relative to $B_X$. Thus for sufficiently
small $\theta_\mui$, polarization alignment will not be achieved for
most emission points.  Figure~\ref{fig:1D_theta_mu5_diffthetar} shows
the 1D profiles of the final $Q$, $U$, $V$ and $\Gamma_{\rm t}$ for a
fixed $\theta_\ri$ and varying $\varphi_\ri$, all with
$\theta_\mui=5^o$. In the left and middle panels where
$\theta_\ri=10^o$ and $20^o$, the profiles are similar to the cases
examined before (see Figs.~\ref{fig:2D_3t} and \ref{fig:1D_B}). For the
$\theta_\ri=30^o$ case (the right panels of
Fig.~\ref{fig:1D_theta_mu5_diffthetar}), $\Gamma_{\rm t}$ is always
less that 3. The reason is that when $\theta_\ri$ is large, the QT
point lies far away from the star $r_{\rm t}\gg R_\ast$ (and it is
even possible that $r_{\rm t}>r_{\rm pl}$ for sufficiently large
$\theta_\ri$). In this case (small $\theta_\mui$ and emission from the
region far away from the magnetic pole), the simple prescription we
have presented in section 4.2~--~4.3 to account for the QT effect
cannot be used, and numerical ray integration from each emission point
is necessary.

\begin{figure*}
\begin{tabular}{c}
\psfig{figure=xray_2D_theta_mu15_holestar.ps,angle=-90,height=10cm} \\
\psfig{figure=xray_2D_theta_mu15_QTEregion.ps,angle=-90,height=4cm} \\
\end{tabular}
\caption{The quasi-tangential effect for X-rays emitted from the whole
  surface of the NS with the $\veck$-$\vecmui$ angle
  $\theta_\mui=15^o$. Here we use the $XYZ$ frame with
  $\vechatk\parallel\vechatZ$, $\vecmui$ in $XZ$-plane and the circle
  corresponds to the project star surface. In the upper panel, the
  hatched region is the effective QT region, defined by $\Gamma_{\rm
    t}<3$.  The bars show the polarization direction. The cross marks
  the projected dipole axis $\vecmui$ on the stellar surface.  The
  lower panel shows the enlarged effective QT region.  The other input
  parameters are: $B_\ast = 10^{14}$\,G, $E =1$\,keV.
\label{fig:2D_theta_mu15}}
\end{figure*}

\begin{figure*}
\begin{tabular}{c}
\psfig{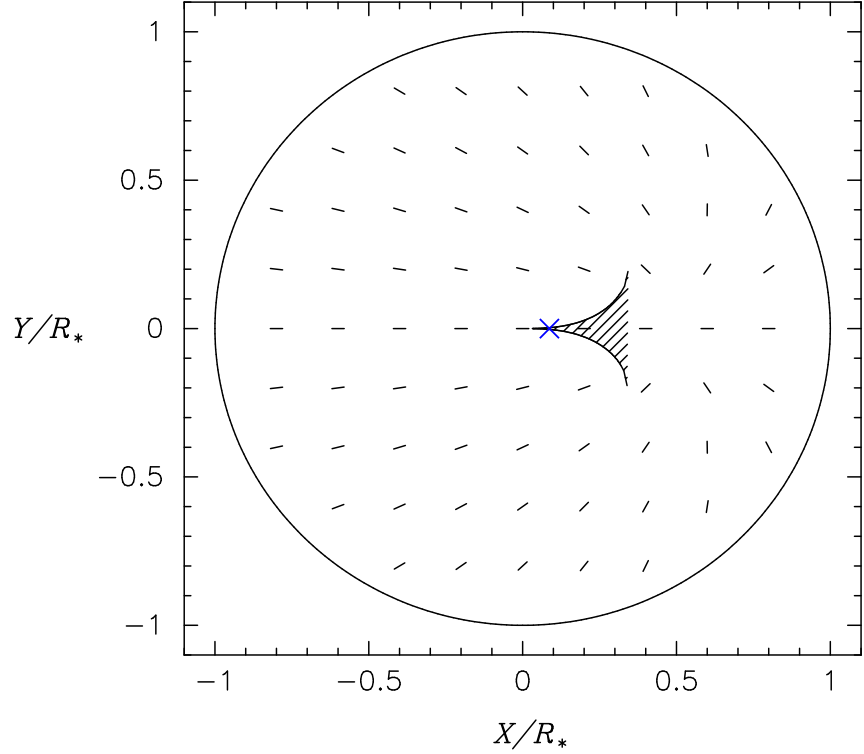} \\
\end{tabular}
\caption{Same as the upper panel of Fig.~\ref{fig:2D_theta_mu15}, except
with $\theta_\mui=5^o$.
\label{fig:2D_theta_mu5}}
\end{figure*}

\begin{figure*}
\begin{tabular}{ccc}
\psfig{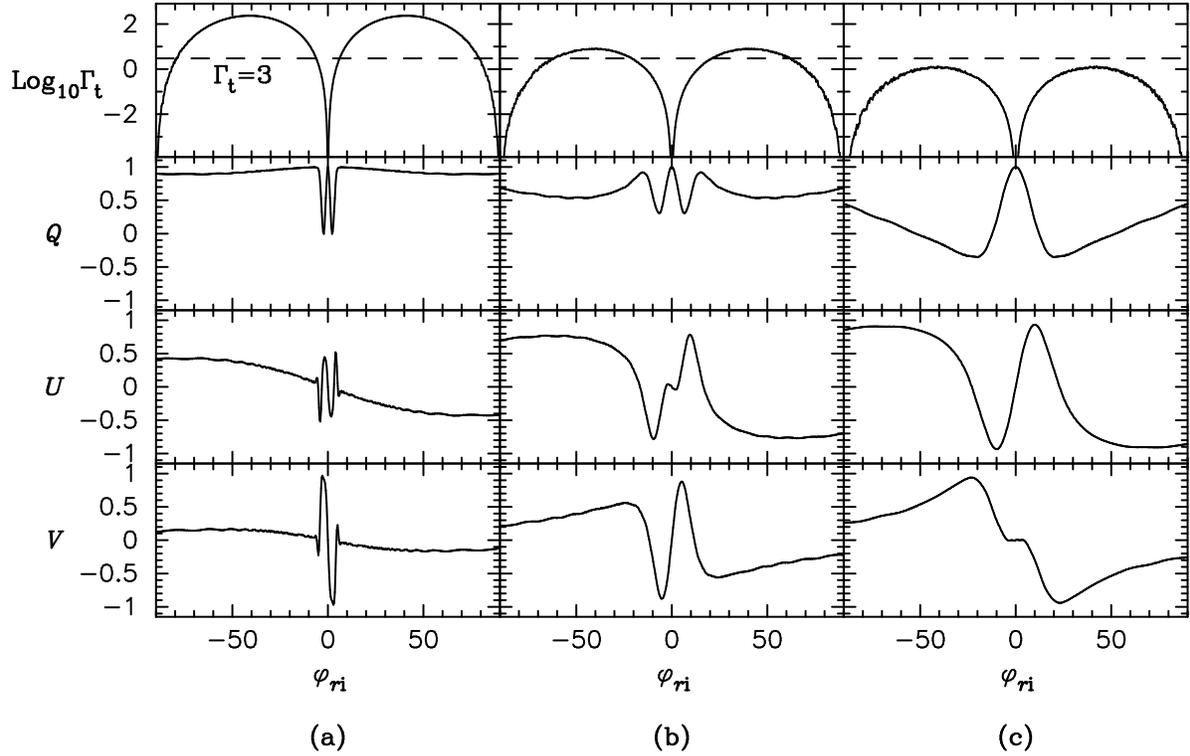} \\
\end{tabular}
\caption{ Similar to Fig.~\ref{fig:2D_3t}, but for emission from
  larger regions of the NS. Here we use the varying $\varphi_\ri$ as
  the $X$-axis. The three columns (from left to right) correspond to a
  fixed $\theta_\mui=10^o,~20^o,~30^o$ (or
  $X/R_\ast=0.174,~0.342,~0.5$ at $Y=0$).  The input parameters are
  $B_\ast=10^{14}\,$G, $E=1$\,keV, $\theta_\mui=5^o$.
  \label{fig:1D_theta_mu5_diffthetar}}
\end{figure*}

\section{Discussion}

We have studied the evolution of photon polarization in a neutron star
magnetosphere whose dielectric property is dominated by vacuum
birefringence. We have focused on X-rays because of the potential of
using X-ray polarimetry to constrain neutron star magnetic fields and
to probe strong-field QED (see Heyl \& Shaviv 2002; Heyl et al.~2003;
Lai \& Ho 2003b).

\subsection{X-ray Polarization Signals without QT Effect}

If one neglects the QT propagation effect studied in this paper, then
it is straightforward to obtain the obserevd polarized X-ray fluxes
(Stokes parameters) from the fluxes at the emission region, at least
approximately, without integrating the polarization evolution
equations in the magnetosphere (Lai \& Ho 2003b; van Adelsberg \& Lai
2006). For a given (small) emission region of projected
area\footnote{This is the area perpendicular to the ray at the
  emission point --- General Relativistic light bending effect can be
  easily included in this.}  $\Delta A_\perp$, one need to know the
intensities of the two photon modes at emission, $I_\perp$ and
$I_\parallel$. In the case of thermal emission, these can be obtained
directly from atmosphere/surface models. As the radiation propagates
through the magnetosphere, the photon mode evolves adiabatically,
following the variation of the magnetic field, until the polarization
limiting radius $r_{\rm pl}$, at which point the polarization is
frozen. Thus the polarized radiation flux beyond $r_{\rm pl}$ is
$F_Q=(I_\parallel-I_\perp)\Delta A_\perp/D$, and $F_U\simeq F_V\simeq
0$\footnote{Note that $F_V$ is not exactly zero because of the neutron
  star rotation and because mode recoupling does not occur instantly
  at $r_{\rm pl}$; see van Adelsberg \& Lai (2006).}, where $D$ is the
distance of the source, $F_Q$ and $~F_U$ are defined in the coordinate
system such that the stellar magnetic field at $r_{\rm pl}$ lies in
the $XZ$ plane (with the $Z$-axis pointing towards the
observer). Since $r_{\rm pl}$ is much larger than the stellar radius,
the magnetic fields as "seen" by different photon rays are aligned and
are determined by the dipole component of the stellar field, one can
simply add up contributions from different surface emission areas to
$F_Q$ to obtain the observed polarization fluxes.

Note that the description of the polarization evolution in the last
paragraph is valid regardless of the possible complexity of the
magnetic field near the stellar surface. This opens up the possibility
of constraining the {\it surface} magnetic field of the neutron star
using X-ray polarimetry.  For example, the polarization light curve
(particularly the dependence on the rotation phase) depends only on
the dipole component of the magnetic field, while the intensity
lightcurve of the same source depends on the surface magnetic
field. On the other hand, the linear polarization spectrum (i.e., its
dependence on the photon energy) depends on the magnetic field at the
emission region (Lai \& Ho 2003b; van Adelsberg 2006); thus it is
possible that a NS with a weak dipole field ($\lo 7\times10^{13}$\,G)
may exhibit X-ray polarization spectrum characteristic of a
$B\go7\times10^{13}$\,G NS (see section~1).

\subsection{Effect of QT Propagation}

The QT propagation effect studied in this paper complicates the above
picture somewhat. When a photon passes through the QT region (which
typically lies within a few stellar radii), its polarization modes can
be temporarily recoupled, and this can give rise to partial mode
conversion. Thus after passing through the QT point, the mode
intensities change to ${\bar I}_\parallel\neq I_\parallel$ and ${\bar
  I}_\perp\neq I_\perp$. The observed polarization flux is then ${\bar
  F}_Q=({\bar I}_\parallel -{\bar I}_\perp)\Delta A/D\neq F_Q$.

In the most general situations, to account for the QT propagation
effect, it is necessary to integrate the polarization evolution
equations (see section 2.2) in order to obtain the observed radiation
Stokes parameters. However, we show in this paper that for generic
near-surface magnetic fields, the effective region where the QT effect
leads to significant polarization changes covers only a small area of
the neutron star surface [see section 3, particularly eq.~(3.22)]. For
a given emission model (and the size of the emission region) and
magnetic field structure, one can use our result in section 3 to
evaluate the importance of the QT effect. In the case of surface
emission from around the polar cap region of a dipole magnetic field,
we have quantified the effect of QT propagation in detail (section 4)
and provided a simple, easy-to-use prescription to account for the QT
effect in determining the observed pollarizatin fluxes. Our key
results are presented in sections 4.2~--~4.3, particularly Fig.~11 and
equations~(4.32) (together with Fig.~10) and (4.35)~--~(4.36). As
discussed in section 4.3, the net effect of QT propagation is to
reduce the degree of linear polarization, so that ${\bar F}_Q/F_Q<1$,
with the reduction factor depending on the photon energy, magnetic
field strength, geometric angles, rotation phase and the emission
area.  The largest reduction is about a factor of two, and occurs for
a particular emission size ($W_{\rm t}/W_{\rm em}\simeq 1.8$; see
Fig.~11). Obviously, for emission from a large area of the stellar
surface, the QT effect is negligible.

Overall, the QT effect will have a small or at most modest effect on
the observed X-ray polarization signals from magnetized neutron stars.
Thus the the description of X-ray polarizaton signals given in section
5.1 remains largely valid.  In most situations (the exceptions are
discussed in section 4.4), the QT effect can be accounted for in a
straightforward manner using the result presented in our paper.

\section*{Acknowledgments}

This work has been supported in part by NASA Grant NNX07AG81G, NSF
grants AST 0707628, and by {\it Chandra} grant TM6-7004X (Smithsonian
Astrophysical Observatory). Chen Wang is also supported by the National
Natural Science Foundation (NNSF) of China (10833003) and the
Initialization Fund for President Award owner of Chinese Academy of
Sciences.


\appendix
\section{Derivation of Equation~(4.31)}

Consider the emission from the polar cap region of a non-rotating NS.
In the $XYZ$-frame (see Fig.~\ref{fig:embeam}, $\veck\parallel
\vechatZ$, $\vecmui$ in $XZ$-plane), the initial magnetic momentum is
\begin{equation}
\vecmui=\mu(\sin\theta_\mui, 0, \cos\theta_\mui),
\end{equation}
and the photon position is
\begin{equation}
\vecr=(R_\ast\sin\theta_\ri, Y, R_\ast\cos\theta_\ri+s),
\end{equation}
where $s$ is the displacement from the emission point, and $Y\ll
R_\ast$.  We want to evaluate $\Gamma_{\rm t}$ at the QT point
$s=s_{\rm t})(Y)$ as a function of $Y$. The magnetic field is given by
eq.~(\ref{eq:B}), and has components:
\begin{eqnarray}
B_X&=&-\frac{\sin\theta_\mui}{r^3}
  +\frac{3R_\ast\sin\theta_\mui}{r^5}(R_\ast+s\cos\theta_\mui),\nonumber\\
B_Y&=&\frac{3Y}{r^5}(R_\ast+s\cos\theta_\mui), \label{eq:B_XY}
\end{eqnarray}
(here we set $\mu=1$). For a given $Y$, The QT point ($s=s_{\rm
  t}(Y)$) is defined by
\begin{equation}
\frac{\rd B_\perp^2}{\rd s}\mid_{r=r_{\rm t}}=0. \label{eq:qtp}
\end{equation}
We can expand $B_X$ and $B_Y$ near $s=s_{\rm t}(Y)$ using Taylor
expansion:
\begin{eqnarray}
B_X&=& B_{X \rm t}+F_1\Delta s\nonumber\\
B_Y&=& (H+F_2\Delta s) Y \label{eq:B_XY_TE}
\end{eqnarray}
with $\Delta s=s-s_{\rm t}$. Note that $B_{X\rm t}$,
$H$, $F_1$, $F_2$ all depend on $s_{\rm t}(Y)$. Substitute
eq.~(\ref{eq:B_XY_TE}) into eq.~(\ref{eq:qtp}), we have
\begin{equation}
B_{X\rm t}=-Y^2\frac{HF_2}{F_1}. \label{eq:BXt}
\end{equation}
From eq.~(\ref{eq:BXt}) we see that 
\begin{equation}
\Delta s_{\rm t}(Y)=s_{\rm t}(0)+{\cal O}(Y^2).
\end{equation}
Thus
\begin{equation}
F_1=F_{10}+{\cal O}(Y^2),~F_2=F_{20}+{\cal O}(Y^2),~H=H_0+{\cal O}(Y^2),
~B_{X\rm t}=-Y^2\frac{H_0F_{20}}{F_{10}}+{\cal O}(Y^4)
\end{equation}
where $F_{10}$, $F_{20}$, $H_0$ are constants.  .  The azimuthal angle
of $\vecB$ is determined by $\tan\varphi_B=B_Y/B_X$. Therefore
$\varphi_B'$ at $s=s_{\rm t}$ is
\begin{eqnarray}
\varphi_B'\mid_{s=s_{\rm t}}&=&\cos^2\varphi_B\left(\frac{B_Y}{B_X}\right)'\mid_{r=r_{\rm t}}
=\frac{YHF_1+Y^3HF_2^2/F_1}{Y^2H^2+Y^4(HF_2/F_1)^2} \propto 1/Y \label{eq:dphi_B}
\end{eqnarray}
We also have
\begin{equation}
\sin\theta_B\mid_{s=s_{\rm t}}=\frac{B_\perp}{B}\propto Y \label{eq:sintheta_B}
\end{equation}
Substitute eq.~(\ref{eq:dphi_B}) and (\ref{eq:sintheta_B}) into
eq.~(\ref{eq:Gamma_ad2}), the adiabaticity parameter at 
QT point is then given by
\begin{equation}
\Gamma_{\rm t}\propto B_{\ast}^2EY^3. \label{eq:GG_Y}
\end{equation}
The QT effective width $W_{\rm t}$, defined by $\Gamma_{\rm t}=3$,
is then $W_{\rm t}\propto (B_{\ast}^2E)^{-1/3}$.

\label{lastpage}

\end{document}